\documentclass[
 aip,
 amsmath,amssymb,
preprint,%
]{revtex4-1}

\usepackage{graphicx}
\usepackage{dcolumn}
\usepackage{bm}

\usepackage[utf8]{inputenc}
\usepackage[T1]{fontenc}
\usepackage{mathptmx}
\usepackage{etoolbox}

\usepackage{xcolor}
\usepackage[normalem]{ulem} 
\usepackage{siunitx}

\makeatletter
\def\@email#1#2{%
 \endgroup
 \patchcmd{\titleblock@produce}
  {\frontmatter@RRAPformat}
  {\frontmatter@RRAPformat{\produce@RRAP{*#1\href{mailto:#2}{#2}}}\frontmatter@RRAPformat}
  {}{}
}%

\begin{document}

\preprint{}

\title{Analytical calculation of the kinetic $q$ factor and resonant response of toroidally confined plasmas}
 
\author{Y. Antonenas}
\affiliation{School of Applied Mathematical and Physical Sciences, National Technical University of Athens, Athens 15780, Greece}

\author{G. Anastassiou}
\affiliation{School of Applied Mathematical and Physical Sciences, National Technical University of Athens, Athens 15780, Greece}

\author{Y. Kominis}
\affiliation{School of Applied Mathematical and Physical Sciences, National Technical University of Athens, Athens 15780, Greece}
\email{gkomin@central.ntua.gr}
\date{\today}
             
\begin{abstract}
Symmetry-breaking perturbations in axisymmetric toroidal plasma configurations have a drastic impact on particle, energy, and momentum transport in fusion devices, thereby affecting their confinement properties. The perturbative modes strongly affect particles with specific kinetic characteristics through resonant mode-particle interactions. In this work we present an analytical calculation of the kinetic $q$ factor enabling the identification of particles with kinetic properties that meet the resonant conditions. This allows us to predict the locations and structures of the corresponding resonant island-chains, as well as the existence of transport barriers in the particle phase space. The analytical results, derived for the case of a large aspect ratio configuration, are systematically compared to numerical simulations, and their domain of validity is thoroughly investigated and explained. Our findings demonstrate that calculating the kinetic
$q$ factor and its dependence on both particle and magnetic field characteristics, provides a valuable tool for understanding and predicting the resonant plasma response to non-axisymmetric perturbations. Moreover, this approach can be semi-analytically applied to generic realistic experimental equilibria, offering a low-computational-cost method for scenario investigations under various multi-scale perturbative modes.
\end{abstract}

\maketitle

\section{Introduction}
The presence of symmetry-breaking perturbations in an axisymmetric MHD equilibrium significantly affects particle, energy and momentum transport, thereby determining the confinement properties and the overall performance of a toroidal fusion device\cite{White1983,Sigmar1992, ITER_Ch2_1999, Heidbrink2020}. Such perturbations may originate either from intrinsic MHD modes such as edge localized modes (ELMs) appearing during H-mode operation in tokamak experiments \cite{Zohm1996,Suttrop2011} and Alfvénic eigenmodes (AEs) \cite{Cheng1986,Chen2007}, or from externally applied modifications to the equilibrium magnetic field, such as Resonant Magnetic Perturbations (RMPs) \cite{Hender1992, Jakubowski2009, Liu2010, Suttrop2011, Spizzo2014, Evans2015}. In all cases, their presence has a drastic effect on single particle orbits as well as on the collective plasma behavior through resonant mode-particle interactions that mostly affect particles with specific drift-orbit frequencies \cite{Xu2018, Todo2019,  He2019, He2020, White2021}. Since orbital frequencies are strongly dependent on the kinetic characteristics of the particles (energy, momentum and pitch angle), the effect of the perturbations is strongly inhomogeneous within the phase space of the particle motion \cite{Kaufman1972, Escande1985}. For small and moderate perturbation strengths, the impact is highly localized in regions where resonance conditions are met and resonant island chains are formed, whereas for larger perturbation strengths, under resonance overlap conditions, the effects extend across broader chaotic regions of the phase space \cite{Lieberman1992}.

A detailed study of particle, energy and momentum transport in the phase space demands significant computational resources for orbit-following simulation codes, in order to track a large number of particles for extended time intervals, under the presence of each specific set of perturbing modes \cite{Shinohara2018, Shinohara2020, Bierwage2022}. These requirements hinder the use of such codes in scenario simulations that necessitate extensive parameter scans, including various sets of modes. Additionally, without a priori knowledge of the phase space regions that are actually affected by each perturbation set, the initial conditions for the simulated particles have to be chosen randomly, leading to a large number of initial conditions corresponding to particle orbits that are essentially unaffected by the specific perturbations, especially in the quite common case where the effects of the perturbations are strongly localized in the phase space. 

 This work aims to present a low-computational-cost calculation capable of providing a priori knoweledge of the exact locations in the phase space where resonant island chains are formed and the number of islands in each chain, as well as the existence and location of transport barriers, crucially affecting the confinement properties of a toroidal fusion device \cite{Falessi2015, Marcus2019, Anastassiou2024}. The calculation refers to the orbital frequency spectrum of the unperturbed guiding center (GC) motion in an axisymmetric equilibrium \cite{Littlejohn83, White1984, Brizard2009}, namely the bounce/transit poloidal frequency and the averaged toroidal precession frequency \cite{WhiteBook}, as well as their ratio, defined as the kinetic $q$ factor, $q_{kin}$\cite{Gobbin2008, Zestanakis2016, Anastassiou2024}. The latter depends on the characteristics of the MHD equilibrium magnetic field (shape and safety factor) as well as on the kinetic characteristics of the particles. It is worth emphasizing that this calculation is performed only once for a specific equilibrium magnetic field and provides predictive capabilities for any set of perturbative modes, suggesting a reduced model and a valuable tool for scenario investigations.  

The kinetic $q$ factor essentially differs from the magnetic $q$ (safety) factor for particles with non-negligible drifts, such as energetic ions resulting in significant differences between the magnetic and the kinetic phase space \cite{White2022a, White2022b, White2023, Moges2024}. For a given axisymmetric equilibrium $q_{kin}$ is a function of the three Constants Of the Motion (COM) of the GC motion, that is the energy, magnetic moment and canonical toroidal momentum \cite{WhiteBook}. Rational values of $q_{kin}$ correspond to resonance conditions \cite{Gobbin2008, Zestanakis2016, Shinohara2018, Shinohara2020}, while its local extrema indicate conditions for the existence of transport barriers \cite{Anastassiou2024}. Both types of conditions can be depicted in the COM particle space along with curves characterizing particles as trapped or passing, and confined or lost, providing a unique overview of the resonant mode-particle interaction. 

The calculation of the kinetic $q$ factor is illustrated for the case of a Large Aspect Ratio (LAR) equilibrium magnetic field [\citenum{WhiteBook}] (p. 98), for which it can be analytically performed. The analytical results are systematically compared and shown in a remarkable agreement with numerical findings, and their domain of validity is investigated and explained in connection to the size of the drift motion and the magnetic safety factor. Both the locations and the number of islands, as well as the locations of transport barriers in the phase space, are accurately predicted by the analytical findings, as further confirmed by numerically obtained Poincaré surfaces of section. Moreover, the presence of multiple island chains, as well as the non-trivial dependence of the number of islands on the  toroidal and poloidal numbers of the perturbing modes, are elucidated and predicted based on the unperturbed motion. It is worth emphasizing that the calculation of kinetic $q$ factor for more general axisymmetric equilibria can be similarly performed in a semi-analytical computationally efficient fashion, enabling the utilization of this valuable tool in cases of realistic experimental equilibria (to be presented in a future paper). 

The paper is organized as follows: In Section \ref{GC System}, the canonical description of the GC motion is discussed, while its formulation in terms of Action-Angle variables is presented in Section \ref{Action-Angle Transformation} for an axisymmetric equilibrium. In Section \ref{Analytical} the kinetic $q$ factor is analytically calculated, systematically compared to numerical results and depicted in the COM space, for several characteristic cases. The resonant response to non-axisymmetric perturbations and the resulting modifications of the phase space as thoroughly examined using appropriate Poincaré surfaces of section along with the predictive accuracy of the analytical results are discussed in Section \ref{Action-Angle Transformation}. The main results and conclusions are summarized in Section \ref{Conclusions}. Additionally, the orbit classification scheme is presented in Appendix A, and the details concerning the relation between the geometrical angles and the Angle variables of the Action-Angle formulation are discussed in Appendix B.  

\section{Canonical description of the Guiding Center motion} \label{GC System}

A general expression of the magnetic field corresponding to a MHD equilibrium that facilitates the Hamiltonian formulation of the GC motion is provided by the utilization of straight field line coordinates. In straight-field line coordinates, such as the Boozer coordinates \cite{Boozer1981, WhiteBook}, the magnetic field is represented in contravariant form as 
\begin{equation}\label{gen mag field} 
    \mathbf{B} = \mathbf{\nabla}\psi\times\mathbf{\nabla}\theta - \mathbf{\nabla}\psi_{p}\times\mathbf{\nabla}\zeta,
\end{equation}
where $\zeta$ and $\theta$ are the toroidal and the poloidal angles, $\psi$ and $\psi_{p}$ are the toroidal and poloidal fluxes of the magnetic field, related through the safety factor $q(\psi) = d\psi/d\psi_{p}$. For the case of an axisymmetric magnetic field configuration, where all partial derivatives with respect to $\zeta$ are equal to zero, the covariant representation of the magnetic field is
\begin{equation}
    \mathbf{B} = g(\psi)\mathbf{\nabla}\zeta+I(\psi)\mathbf{\nabla}\theta + \delta(\psi,\theta)\mathbf{\nabla}\psi
\end{equation}
where the functions $g(\psi)$, $I(\psi)$ are related to the poloidal and toroidal current, and $\delta(\psi,\theta)$ measures the non-orthogonality of the Boozer coordinate system.

The GC motion of a charged particle in a  magnetic field is described by the Lagrangian \cite{Littlejohn83},
$L = (\mathbf{A}+\rho_{||}\mathbf{B}) \cdot \mathbf{u}+\mu\dot{\xi}-H$, where $\mathbf{A}$ and $\mathbf{B}$ are the vector potential and the magnetic field, $\mathbf{u}$ is the GC velocity, $\mu$ is the magnetic moment, $\xi$ is the gyro-phase, $\rho_{||}$ is the velocity component parallel to the magnetic field, normalized to B, and 
\begin{equation}\label{GC H}
    H = \frac{\rho_{||}^2}{2} B^2 + \mu B
\end{equation}
is the Hamiltonian, in the absence of an electrostatic potential, corresponding to the particle energy $E$. The GC motion is described in normalized units where distance is normalized to the major radius $R_0$, and time is normalized to $\omega_{0}^{-1}$ , with $\omega_{0}=qB_{0}/m$ the on-axis gyro-frequency; hence for different particle species, the GC drift velocities scale with the mass-to-charge ratio. Energy is normalized to $m\omega_{0}^2R^{2}$ and the magnetic field is normalized to its on-axis value $B_{0}$ \cite{Littlejohn83, WhiteBook}.  The gyro-radius is $\rho=u_{\bot}/B<<1$ and the magnetic moment $\mu=u_{\bot}^2/(2B)$, as well as the cross-field drift, are of order $\rho^2$.

The canonical momenta conjugate to the coordinates $\theta$ and $\zeta$ are expressed in terms of the $\rho_{||}$ and $\psi$ \cite{White1984} as 
\begin{align}\label{canon moments}
    P_{\theta} = \psi + \rho_{||} I(\psi), \qquad  P_{\zeta} = \rho_{||} g(\psi) - \psi_{p}(\psi),
\end{align}
while $\mu$ is the canonical momentum conjugate to the gyro-angle $\xi$. Consequently, the Hamiltonian \eqref{GC H} can be written as
\begin{equation} \label{GC H 1}
    H = \frac{\left[P_{\zeta} + \psi_{p}(P_{\zeta},P_{\theta})\right]^2}{2 g^2(P_{\zeta},P_{\theta})} B^2(P_{\zeta}, P_{\theta}, \theta) + \mu B(P_{\zeta}, P_{\theta}, \theta),
\end{equation}

Since the Hamiltonian does not explicitly depend on time, the energy $E$ of the system is a constant of the motion. Moreover, the absence of the angles $\zeta$ (due to axisymmetry) and $\xi$ (due to gyro-averaging) in the Hamiltonian \eqref{GC H 1} results in the constancy of their conjugate momenta, $P_{\zeta}$ and $\mu$, respectively. Therefore, the Hamiltonian system is integrable and its orbits can be described in terms of the three constants of the motion $(E, \mu, P_{\zeta})$. Each point in this three-dimensional Constants-Of-the-Motion (COM) space uniquely labels a GC orbit, with the shape of each orbit, as projected in the poloidal plane $(\theta,P_\theta)$, being obtained as a level curve of the Hamiltonian (\ref{GC H 1}). Orbits can  be classified with respect to being trapped or passing (circulating), and confined or lost [\citenum{WhiteBook}] (Sec. 3.3), depending on their location relatively to well-defined curves in the COM space (see Appendix A). The corresponding diagrams provide a valuable tool for the study of the orbital spectrum and the impact of resonances on particle transport as will be demonstrated in the following sections.

\section{Action-Angle formulation} \label{Action-Angle Transformation}
The GC Hamiltonian  can be formulated in Action-Angle (AA) variables through a canonical transformation. The AA formulation provides unique advantages for the study of complex particle dynamics, by exploiting the full range of concepts and methods of Hamiltonian systems. When expressed in AA variables, the different time scales of multiply periodic particle dynamics are distinctly separated into different degrees of freedom. The respective orbital frequencies can be readily calculated, without requiring a complete solution to the equations of the motion, facilitating the formulation of resonance conditions that dictate particle, momentum and energy transport, under the influence of external perturbations to the integrable axisymmetric configuration. Moreover, the AA variables offer a powerful means for a systematic dynamical contraction of the GC dynamics in a hierarchy of descriptions with reduced dimensionality \cite{Goldstein, WhiteBook, Brizard2014, YA}.

For the GC Hamiltonian (\ref{GC H 1}) the three Action variables are defined as 
\begin{equation} \label{gen Action Integral}
    J_{i} = \frac{1}{2\pi}\oint {p_{i}(q_{i}; E,\mu,P_\zeta)dq_{i}}
\end{equation}
where $(q_i,p_i)=(\xi,\mu), (\zeta,P_\zeta), (\theta,P_\theta)$. The integration is performed over a complete cycle of the respective canonical position $q_i$, with the variables describing the other degrees of freedom kept constant \cite{Goldstein}. For the first two degrees of freedom, the canonical positions $\xi$ and $\zeta$ are cyclic, completing a full circle within the range of $0$ to $2\pi$; their respective canonical momenta $\mu$ and $P_\zeta$ are invariants, and the integration simply yields
\begin{equation}
    J_\xi= \mu, J_\zeta= \sigma P_\zeta
\end{equation}
with $\sigma =+1$ and $\sigma=-1$ for co- and counter-moving orbits, with respect to the magnetic field, respectively. For the third (poloidal) degree of freedom we have
\begin{equation} \label{gen Jtheta}
    J_{\theta} = \frac{1}{2\pi}\oint{P_{\theta}(\theta^{\prime}; E, \mu, P_{\zeta})}d\theta^{\prime},
\end{equation}
where $P_\theta$ is obtained from Eq. (\ref{GC H 1}) as a function of $\theta$ and the three COM. It is worth noting that the explicit appearance of the (non-cyclic) canonical position $\theta$ in the Hamiltonian (\ref{GC H 1}) results in its non-trivial variation; consequently, $\theta$ may be restricted to vary in subsets of the interval $(0,2\pi)$ or may even be stationary. The canonical transformation to the AA variables is provided by the mixed-variable generating function
\begin{equation} \label{genarating function}
   F(\theta, \zeta, \xi, J_{\theta}, J_{\zeta}, J_{\xi}) = J_{\zeta}\zeta + J_{\xi}\xi+ f(\theta,J_{\theta}, J_{\zeta}, J_{\xi})
\end{equation}
where 
\begin{equation}
    f(\theta,J_{\theta}, J_{\zeta}, J_{\xi})=\int^{\theta}{P_{\theta}(\theta^{\prime}, E, \mu, P_{\zeta})}d\theta^{\prime} 
\end{equation}
and the three COM, $(E,\mu,P_\zeta)$, are now considered as functions of the three Actions $\mathbf{J}=(J_\theta,J_\zeta,J_\xi)$.

The new angles are related to the old ones through the canonical transformation as \cite{Kaufman1972, Zestanakis2016}
\begin{equation}\label{eq:Angles}
    \hat{\theta}=\frac{\partial f(\mathbf{J,\theta})}{\partial J_\theta}, \qquad \hat{\zeta}=\zeta+\frac{\partial f(\mathbf{J,\theta})}{\partial J_\zeta}, \qquad \hat{\xi}=\xi+\frac{\partial f(\mathbf{J,\theta})}{\partial J_\xi} . 
\end{equation}
Expressed in AA variables, the Hamiltonian is independent of all new Angles, and the orbital frequencies are given as $\hat{\omega}_i=\partial H(\mathbf{J})/\partial J_i$, $i=\theta,\zeta,\xi$. It is worth emphasizing that, in contrast to the new Angles $(\hat{\theta},\hat{\zeta},\hat{\xi})$, the original angles $(\theta,\zeta,\xi)$ do not have a linear time dependence, and therefore, their time derivatives $(\dot{\theta},\dot{\zeta},\dot{\xi})$ do not correspond to any orbital frequency. The first of Eqs. (\ref{eq:Angles}) implies that the time dependence of the Angle $\hat{\theta}$ is solely through the old angle $\theta$, so that it is also a periodic function of time with the same frequency
\begin{equation}
\omega_\theta=\hat{\omega}_\theta=\frac{\partial H(\mathbf{J})}{\partial J_\theta}.    
\end{equation}
It is also worth noting that $\hat{\theta}$ corresponds to an unbounded phase coordinate with a linear time dependence, whereas $\theta$ can be either unbounded (for passing orbits) or bounded (for trapped orbits) and exhibits a more complex time dependence. The case of the other two Angles, $\hat{\zeta}$ and $\hat{\xi}$ is qualitatively different, since they depend, not only on the corresponding original angles, $\zeta$ and $\xi$, but also on $\theta$ (see also Appendix B). 

By implicit differentiation it can be readily shown \cite{YA, Anastassiou2024} that 
\begin{equation}\label{gen hat omega zeta}
    \hat{\omega}_{\zeta} = \frac{\partial H(\mathbf{J})}{\partial J_\zeta}=-\frac{\partial H}{\partial J_{\theta}}\frac{\partial J_{\theta}}{\partial J_{\zeta}} = -\hat{\omega}_{\theta}\frac{\partial J_{\theta}}{\partial J_{\zeta}}
\end{equation}
and, similarly,
\begin{equation}\label{gen hat omega xi}
    \hat{\omega}_{\xi} =\frac{\partial H(\mathbf{J})}{\partial J_\xi} = -\frac{\partial H}{\partial J_{\theta}}\frac{\partial J_{\theta}}{\partial J_{\xi}} = -\hat{\omega}_{\theta}\frac{\partial J_{\theta}}{\partial J_{\xi}}.
\end{equation}
The form of these equations suggests that $(-J_\theta)$ acts as a new Hamiltonian in the remaining Action-Angle variables, where the new time variable is measured in units of the bounce/transit period, similarly to the definition of the dynamical contraction discussed in [\citenum{WhiteBook}] (p. 97).\

Focusing on the drift motion in the slower degrees of freedom $(\hat{\theta},\hat{\zeta})$, we have  
\begin{equation} \label{precession}
 \hat{\omega}_{\zeta} = -\hat{\omega}_{\theta} \frac{1}{2\pi}\oint{\frac{\partial P_{\theta}(\theta', E, P_{\zeta}, \mu)}{{\partial P_{\zeta}}}} d \theta'= \frac{(\Delta \zeta)_{T_{\theta}}}{T_\theta},
\end{equation}
where we have used a change of variables from $\theta$ to $\zeta$, with $d\zeta / d\theta= - \partial P_{\theta}/\partial P_{\zeta}$, and $(\Delta \zeta)_{T_\theta}$ denotes the variation of $\zeta$ over a complete period $T_\theta$ of the poloidal angle $\theta$, indicating that the orbital frequency $\hat{\omega}_\zeta$ corresponds to the bounce/transit averaged rate of toroidal precession. Similarly, it can be shown that $\hat{\omega}_\xi$ corresponds to the bounce/transit averaged gyrofrequency. Thus, the complete orbital spectrum of the GC motion is determined in terms of the three Action variables $\mathbf{J}$ or the three COM $(E,\mu,P_\zeta)$. It is worth clarifying that in the following we do not consider high-frequency perturbations, introducing $\xi$ dependence, so that the non-axisymmetrically perturbed Hamiltonian system can be treated as a one-parameter $\mu$ family of two-degree of freedom Hamiltonian systems; however, we have preferred to keep a general formulation demonstrating the generality of this approach and suggesting possible extensions to high-frequency perturbations.

\section{Analytical calculation of the kinetic $q$ factor} \label{Analytical}
To calculate the Action $J_{\theta}$, as defined by Eq. \eqref{gen Jtheta}, it is necessary to solve Eq. \eqref{GC H 1} in order to obtain $P_\theta$ as a function of $\theta$ and the three COM $(E,\mu,P_\zeta)$, defining each orbit along which the integration is performed. Even for the simplest case of a LAR equilibrium, such an analytical expression is not always available, and even if available, the subsequent integration cannot be performed analytically. To overcome this issue, we resort to an approximation  by selecting for each GC orbit an appropriate magnetic surface of reference $\psi_0$, where the magnetic field is calculated as $B(\psi_0,\theta)$. Such an approximation is analogous to the GC approximation where the magnetic field at each position of a gyrating particle is taken equal to its value at the center of its orbit. Just as the validity of the GC approximation depends on the spatial variation of the magnetic field within a gyration orbit, the validity of this approximation depends on the variation of the magnetic field within a drift GC orbit, with larger drifts making the approximation less reliable.

For the LAR equilibrium (Appendix A), where $\psi=P_\theta$, the appropriate magnetic surface of reference can be defined as a function of the constants of the motion, $P_{\theta_0}(E,\mu,P_\zeta)$ according to 
\begin{equation} \label{def of Ptheta0}
    P_{\theta_{0}} = \left\{
    \begin{array}{ll}
     \psi_{p}^{-1}(-P_\zeta)    & \text{, trapped}\\
     \psi_{p}^{-1}(\pm \sqrt{2(E-\mu)}-P_\zeta)   & \text{, co/counter-passing}
    \end{array}
    \right.
\end{equation}
where $\psi_0 (P_{\theta_0})$ corresponds to the flux surface at the banana tip $(\rho_\parallel=0)$ and the point $\theta=\pi/2$ of the trapped and passing orbits, respectively \cite{Pinches1998}, and $\psi_p^{-1}$ denotes the inverse function.
Moreover, $\psi_{p}(P_{\theta})$ can be approximated by a first order Taylor expansion around $P_{\theta_0}$, as
\begin{equation}
    \psi_{p}(P_{\theta})=\psi_{p}(P_{\theta_{0}})+\frac{P_{\theta}-P_{\theta_{0}}}{q(P_{\theta_{0}})}.     
\end{equation}
    
The Action can be calculated as follows:

\begin{equation} \label{jt}
J_{\theta} = \frac{q(P_{\theta_{0}})}{2\pi}\oint{\frac{\sqrt{2\left[E - \mu\left(1-\sqrt{2 P_{\theta_{0}}}\cos\theta\right)\right]}}{1-\sqrt{2 P_{\theta_{0}}}\cos\theta}}
      -\left[q(P_{\theta_{0}})\rho_{\parallel 0}(P_\zeta,P_{\theta_0})-P_{\theta_{0}}\right]d\theta
\end{equation}
with 
\begin{equation}
    \rho_{\parallel 0}(P_\zeta,P_{\theta_0})=P_{\zeta} + \psi_{p}(P_{\theta_{0}}),
\end{equation}

The corresponding analytical expressions for trapped $(t)$ and passing $(p)$ particles are respectively given by

\begin{equation}\label{Jtheta_t}
    J^{(t)}_{\theta} =   q(P_{\theta_{0}})\frac{8\sqrt{\mu r}}{\pi\eta(1-r)}\left[(\eta k -1)\Pi(\eta k, k)+K(k)\right]
\end{equation}
\begin{equation}\label{Jtheta_p}
    J^{(p)}_{\theta} = q(P_{\theta_{0}})\frac{4\sqrt{\mu r}}{\pi\eta(1-r)}\left[\frac{\eta k -1}{\sqrt{k}}\Pi(\eta, \frac{1}{k})+\frac{K(\frac{1}{k})}{\sqrt{k}}\right]
            - \sigma\left[q(P_{\theta_{0}})\rho_{\parallel 0}(P_\zeta,P_{\theta_0})-P_{\theta_{0}}\right],
\end{equation}
where $r=\sqrt{2P_{\theta_{0}}}$, $\eta=-2r/(1-r)$, $k = \left(E-\mu(1-r)\right)/2\mu r$ is the trapping parameter with $0<k<1$ for trapped and $k>1$ for the passing orbits, $K,\Pi$ are the complete elliptic integrals of the first and third kind, respectively, while $\sigma =+1$ for trapped and co-passing orbits and $\sigma = -1$ for counter-passing orbits \cite{YA}.

Based on Eqs. \eqref{gen hat omega zeta}, \eqref{Jtheta_t}, \eqref{Jtheta_p}, the kinetic $q$ factor $q_{kin}$ can be analytically calculated (the resulting expression is quite lengthy and therefore ommited) as a function of the three COM as
\begin{equation} \label{q-kinetic}
    q_{kin}(E, \mu, P_{\zeta})\equiv \frac{\hat{\omega}_\zeta} {\hat{\omega}_\theta}
    =-\sigma\cdot\frac{\partial J_\theta^{(t,p)}(E, \mu, P_{\zeta})}{\partial  P_\zeta}.
\end{equation}

\begin{figure}[h!]
    \centering
    \includegraphics[width=0.49\textwidth]{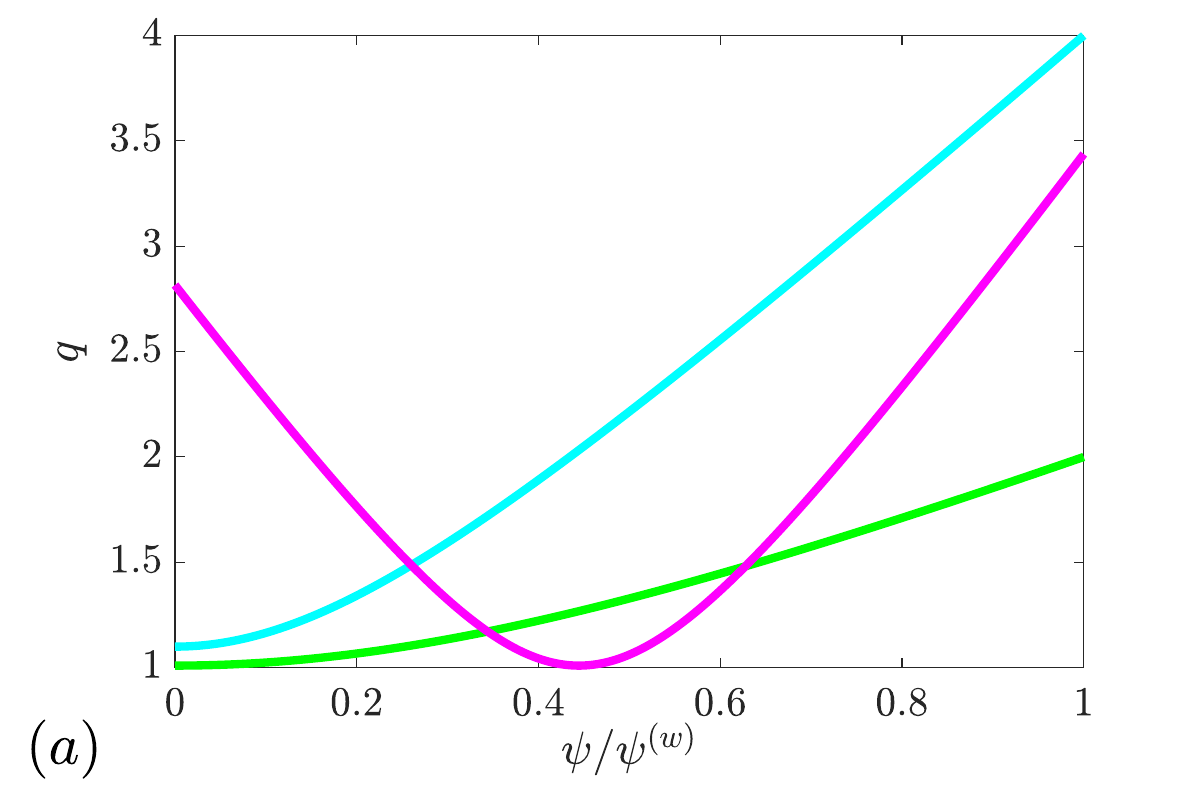}
    \centering
    \includegraphics[width=0.49\textwidth]{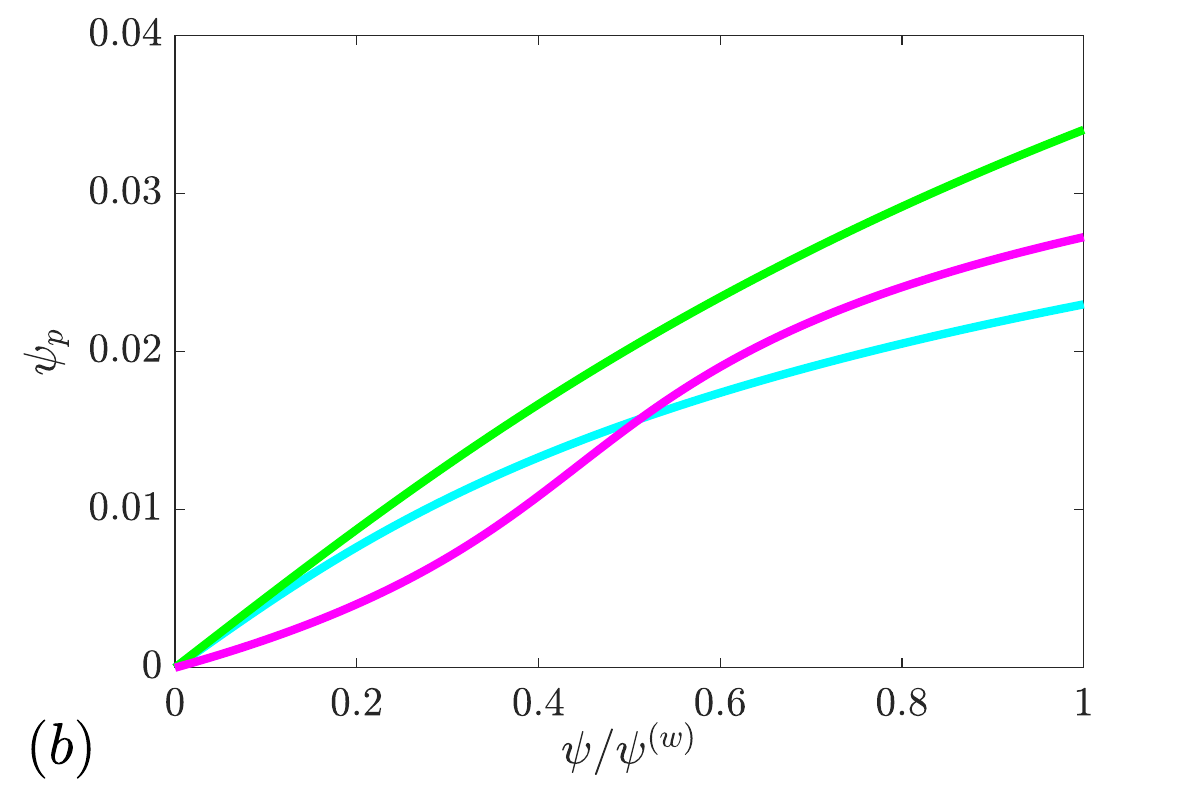}
    \caption{(a): Three different $q(\psi)$ profiles.  All profiles are described by Eq. \eqref{gen q factor} with $\nu=2$. Cyan lines ($q_{1}$):  $q_{a}=1.1$, $q_{w}=4.0$, $\lambda=0$. Green lines ($q_{2}$): $q_{a}=1.01$, $q_{w}=3.0$, $\lambda=0$. Magenta lines ($q_{3}$): $q_{a}=1.01$, $q_{w}=3.0$, $\lambda=0.44$. (b): The corresponding poloidal flux $\psi_{p}(\psi)$ as obtained by integrating each $q$ profile over $\psi$.}
    \label{fig:Fig1}
\end{figure}

In the following, we present analytical calculations of the $q_{kin}$ based on Eqs. \eqref{Jtheta_t}, \eqref{Jtheta_p}, \eqref{q-kinetic} for particles with various kinetic characteristics and drift orbit widths, we investigate its dependence on the magnetic $q$ profile, and systematically compare with numerical results in order to estimate the range of validity of the analytical expressions for a LAR axisymmetric equilibrium. The magnetic $q$ profile is given by the expression [\citenum{WhiteBook}] (Sec. 2.9)
\begin{equation} \label{gen q factor}
    q(\psi) = q_{0} \Bigg[1 + \bigg(\bigg(\frac{q_{w}}{q_{a}}\bigg)^\nu - 1\bigg)(\psi/\psi_{w} - \lambda)^\nu \Bigg]^{1/\nu},
\end{equation}
where $\lambda$ designates the location of the local minimum of the $q$ profile at $\psi/\psi_{w}=\lambda$, corresponding to a flux surface with zero magnetic shear. When $\lambda=0$, the minimum occurs at the magnetic axis $\psi/\psi_{w} = 0$, and $q_a$ and $q_w$ represent the values of $q$ at the magnetic axis and the wall, respectively. Moreover, $\psi_w$ is the toroidal flux at the wall, and $\nu$ controls the radial shape of the $q$-profile.
Three different $q$ profiles (with $\nu=2$) are illustrated in Fig. \ref{fig:Fig1}(a). Both $q_1$ and $q_2$ are monotonic profiles $(\lambda=0)$. The $q_{1}$ profile (cyan line) corresponds to parameters $q_{a}=1.1$, $q_{w}=4.0$, while the $q_{2}$ profile (green line) has a smaller value at the magnetic axis $(q_{a}=1.01)$ and the wall $(q_{w}=3.0$). The $q_{3}$ profile (magenta line) is non-monotonic, and is defined by $q_{a}=1.01$, $q_{w}=6.0$, and $\lambda=0.44$, with its local minimum (at $\psi/\psi_w=\lambda=0.44$) corresponding to a flux surface with zero magnetic shear (note that for this profile $q_a$ and $q_w$ do not correspond to the values of $q$ at the axis and the wall). This profile is an example of a reversed shear magnetic $q$ profile, similar to those created in  Tokamak Fusion Test Reactor (TFTR) experiments, where they have been shown to lead to reduced particle transport \cite{Levinton1995}. Reversed shear configurations have also been shown to be related to the formation of internal transport barriers and reduced transport in DIII-D discharges \cite{Strait1995}, and are expected to reduce alpha particles losses  in ITER \cite{Fasoli2007}.

Evidently, the poloidal flux for each $q$ profile has different dependence on the toroidal flux $\psi_p(\psi)$, which results in a different value at the wall $\psi_{p}^{(w)}$, as depicted in Fig. \ref{fig:Fig1}(b).

The role of the kinetic characteristics of the particles along with the magnetic $q$ profile with respect to the form of the $q_{kin}$ profile, and the validity of the analytical results can be investigated by systematically dissecting the three dimensional COM space. In the following we consider a LAR configuration with major radius $R_0=\SI{1.65}{\metre}$, inverse aspect ratio $r/R_0 = 0.18$ and on-axis magnetic field $B_{0}=\SI{1}{\tesla}$, and investigate hydrogen orbits (atomic number $Z=1$), focusing on four characteristic cases.  

\textbf{Case \#1:} Low-energy particles, with $\mu B_{0} = \SI{2}{\kilo\electronvolt}$ ($\mu = 7.7\times 10^{-6}$ in normalized units), in the $q_1$ profile.  

\textbf{Case \#2:} Mildly energetic particles (in comparison to Case \#1), with $\mu B_{0} = \SI{10}{\kilo\electronvolt}$ ($\mu =3.8\times 10^{-5}$ in normalized units), in the $q_1$ profile.

\textbf{Case \#3:} Mildly energetic particles with the same characteristics as in Case \#2 ($\mu B_{0}=\SI{10}{\kilo\electronvolt}$), in the $q_{2}$ profile, which results in a higher value of the poloidal flux at the wall $\psi_{p}^{(w)}$. 

\textbf{Case \#4:} Mildly energetic particles with the same characteristics ($\mu B_0=\SI{10}{\kilo\electronvolt}$) as in Cases \#2 and \#3, in the non-monotonic $q_3$ profile. This case favors the formation of transport barriers due to the existence of local minima of the kinetic $q$ factor in the COM space. 

\begin{figure}[h!]
    \centering
    \includegraphics[width=0.49\textwidth]{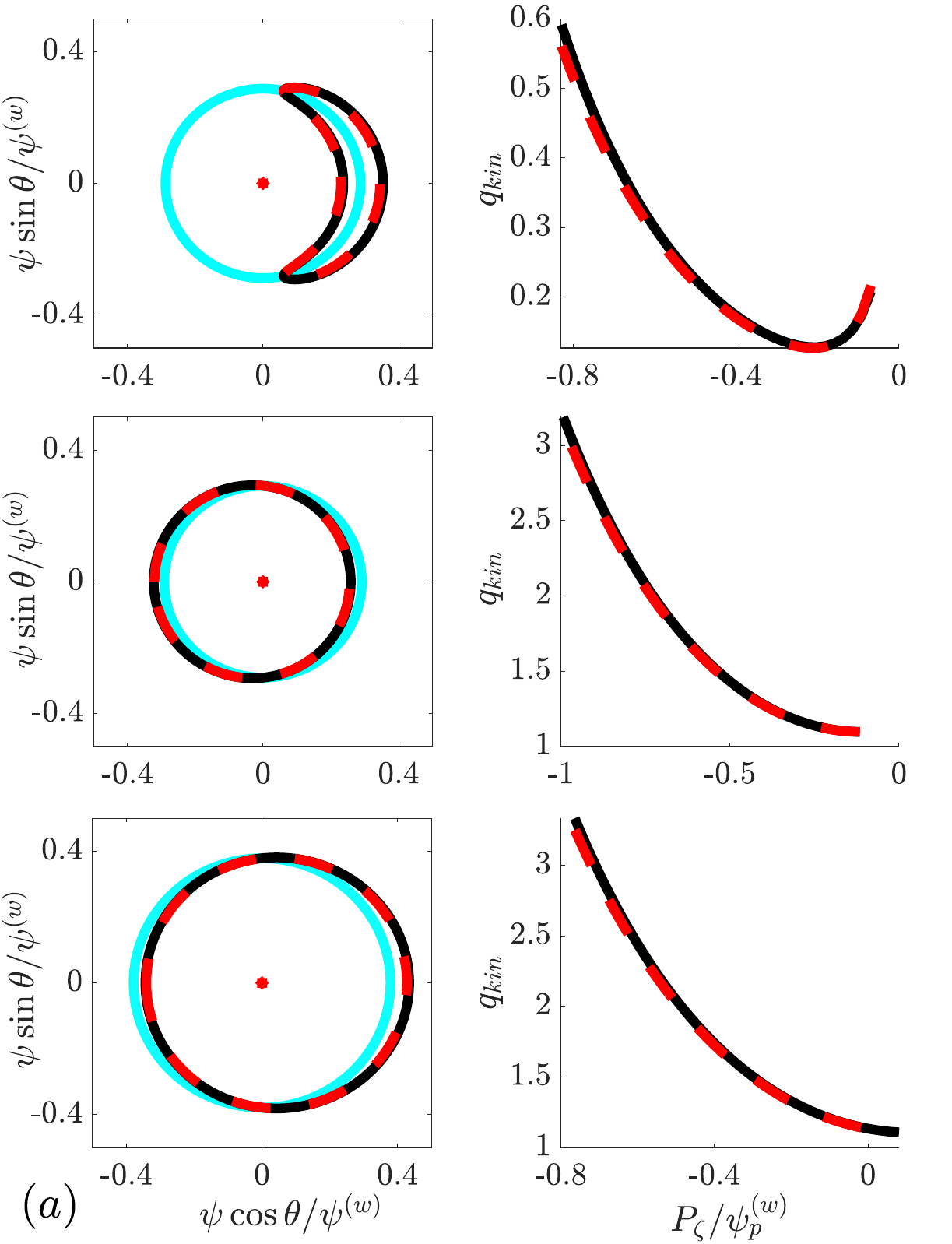}
    \centering
    \includegraphics[width=0.49\textwidth]{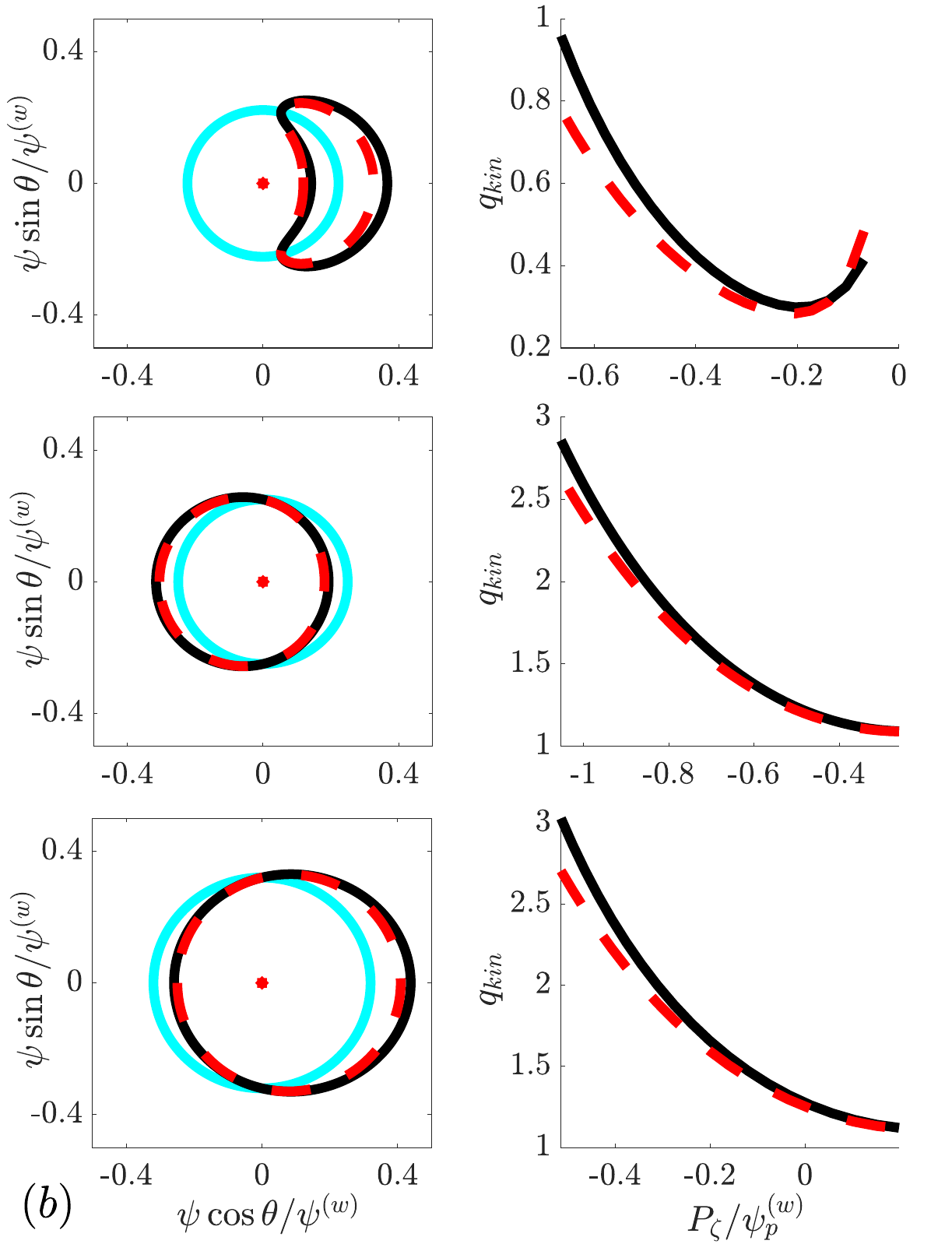}
    \caption{(a) Case \#1: Low-energy particles, with $\mu B_{0} = \SI{2}{\kilo\electronvolt}$, in the $q_1$ profile. (b) Case \#2: More energetic particles with $\mu B_{0} = \SI{10}{\kilo\electronvolt}$, in the $q_1$ profile. From top to bottom: trapped particles with  $E/\mu B_{0} = 0.996$, counter passing, and co-passing particles with  $E/\mu B_{0} = 1.4$. 
    Left columns: characteristic orbits in the configuration space along with their corresponding flux surface of reference $\psi_0=P_{\theta_0}$ (cyan lines).  Right columns: $q_{kin}$ as a function of the canonical momentum $P_\zeta$. Black-solid and red-dashed lines denote numerical and analytical results, respectively. The deviation between analytical and numerical calculations increases with the particle energy and the drift-orbit width.} 
    \label{fig:Fig2}
\end{figure}

\begin{figure}[h!]
    \centering
    \includegraphics[width=0.49\textwidth]{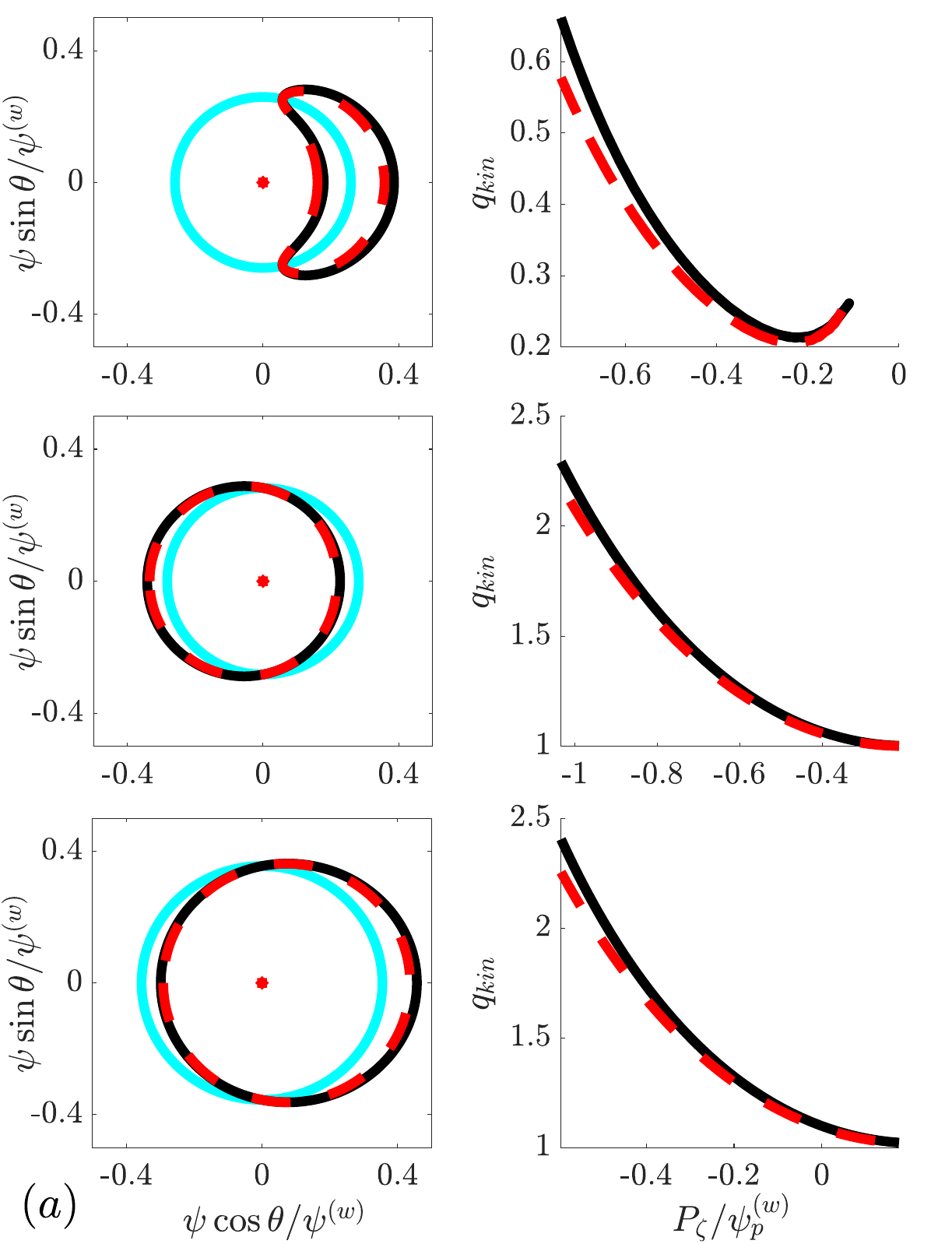}
    \centering
    \includegraphics[width=0.49\textwidth]{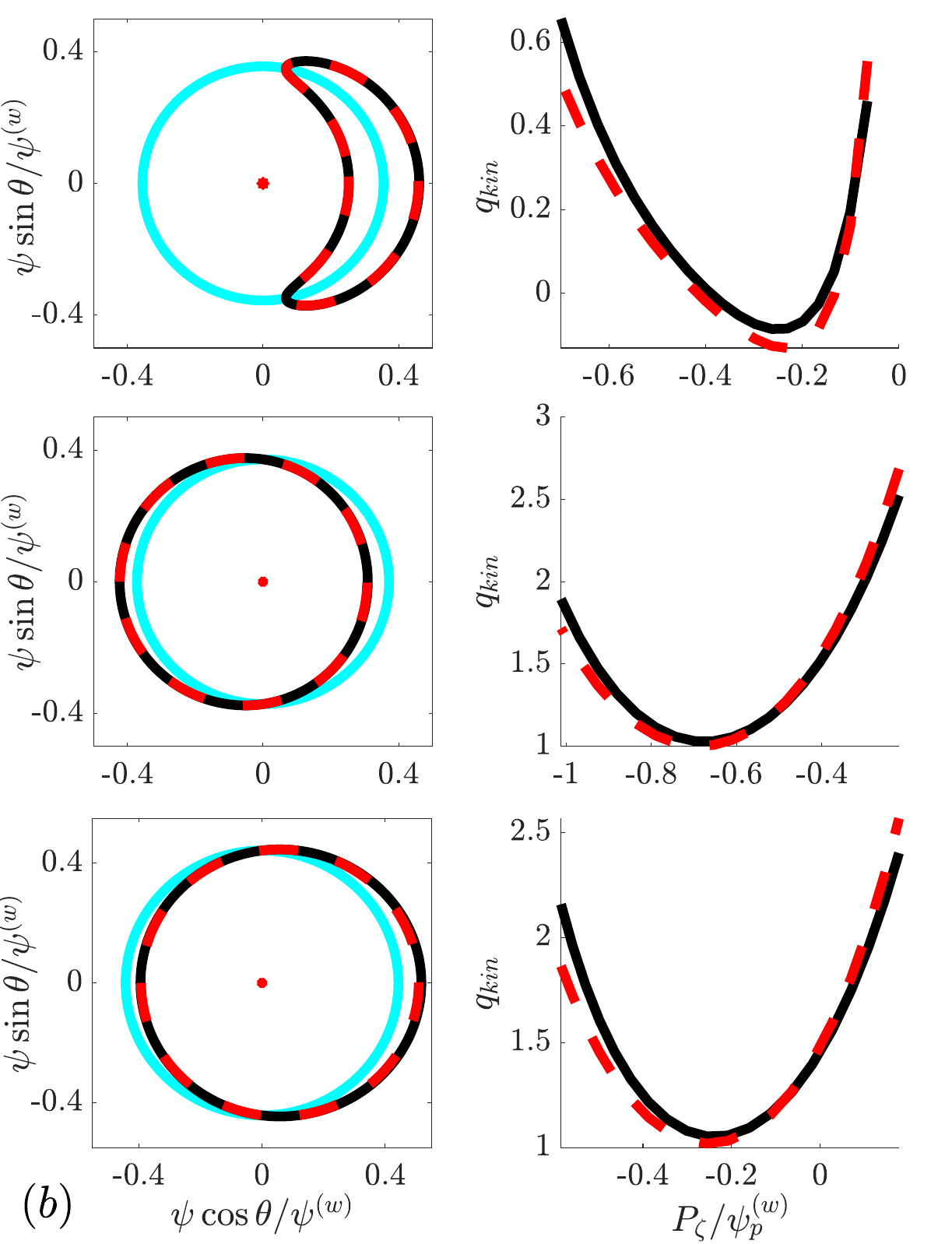}
    \caption{(a) Case \#3: Energetic particles with $\mu B_{0} = \SI{10}{\kilo\electronvolt}$ in the $q_2$ profile. (b) Case \#4: Energetic particles with $\mu B_{0} = \SI{10}{\kilo\electronvolt}$ in the $q_3$ profile.
    From top to bottom: trapped particles with $E/\mu B_{0}= 0.996$, counter passing, and co-passing particles with  $E /\mu B_{0} = 1.4$. 
    Left columns: characteristic orbits in the configuration space along with their corresponding flux surface of reference $\psi_0=P_{\theta_0}$ (cyan lines).  Right columns: $q_{kin}$ as a function of the canonical momentum $P_\zeta$. Black-solid and red-dashed lines denote numerical and analytical results, respectively.
    In Case \#3, the deviations between analytical and numerical calculations are reduced in comparison to Case \#2. In Case \#4, in contrast to the previous cases, a local minimum in $q_{kin}$ is observed for passing particles.} 
    \label{fig:Fig3}
\end{figure}

In Figs. \ref{fig:Fig2} and \ref{fig:Fig3}, the analytical calculation of $q_{kin}$ (red dashed lines), based on Eq. \eqref{q-kinetic}, is compared to calculations obtained from the numerical integration of the GC equations of motion (black solid lines), for the four cases. Each orbit is uniquely labeled by a specific set of the COM $(E, \mu, P_{\zeta})$. For a given value of $\mu$, particles with different energies $E$ can be trapped (top), counter-passing (middle), or co-passing (bottom). The first columns depict the poloidal projection of characteristic GC orbits in the configuration space. The analytically obtained orbits (red-dashed lines) are calculated as level sets of the Hamiltonian [Eq. \eqref{GC H 1}] with the magnetic field evaluated at the flux surface of reference $\psi_0=P_{\theta_0}$, given by Eq. \eqref{def of Ptheta0} and depicted by cyan lines. The second columns depict the kinetic $q$ factor as a function of the canonical momentum $P_\zeta$ for trapped and counter/co-passing particles. 
Figures Figs. \ref{fig:Fig2}, \ref{fig:Fig3} clearly show that there is a direct relation between the drift width of a GC orbit and the discrepancy between the analytically and the numerically calculated values of the kinetic $q$ factor, which is based on approximating the value of the magnetic field in the course of a GC orbit with a single value on a flux surface of reference $\psi_0$.  For low-energy particles, the agreement between analytical and numerical calculations is excellent, with maximum deviations of less than $5\%$ and $2\%$ for trapped and passing particles, respectively, as shown for the Case \#1 in Fig. \ref{fig:Fig2}(a). Under the same magnetic field configuration (same magnetic $q$ factor, $q_1$), the accuracy deteriorates for more energetic particles, as shown for the Case \#2 in Fig. \ref{fig:Fig2}(b). However, for the same energetic particles, the accuracy significantly improves when the profile of the magnetic $q$ factor changes from $q_1$ to $q_2$, as depicted in Fig.  \ref{fig:Fig3}(a). It is worth noting that, in all cases, the discrepancies are larger for trapped particles. The case of the non-monotonic $q$ profile ($q_3$) is illustrated in Fig. \ref{fig:Fig3}(b), where a local minimum of $q_{kin}$ is shown to occur for passing particles as well, in contrast to the previous cases where such a local minimum occurred only for trapped particles. As will be shown in the following section, a local extremum of $q_{kin}$ indicates the formation of a transport barrier under the presence of non-axisymmetric perturbations. In this case of passing particles, the minimum occurs for particle orbits located closer to the wall in contrast to the case of trapped particles, where the minimum occurs for particle orbits close to the magnetic axis. 

\begin{figure}[h!]
    \centering
    \includegraphics[width=0.49\textwidth,keepaspectratio]{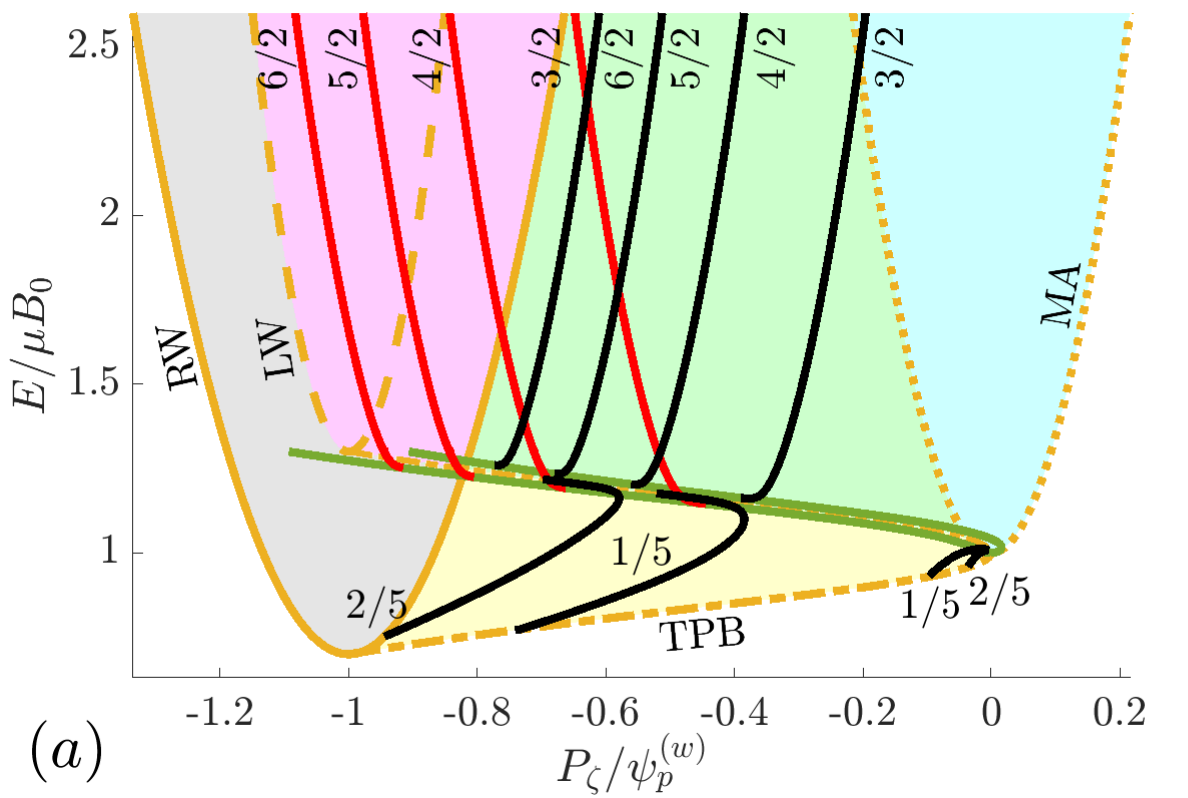}
    \includegraphics[width=0.49\textwidth,keepaspectratio]{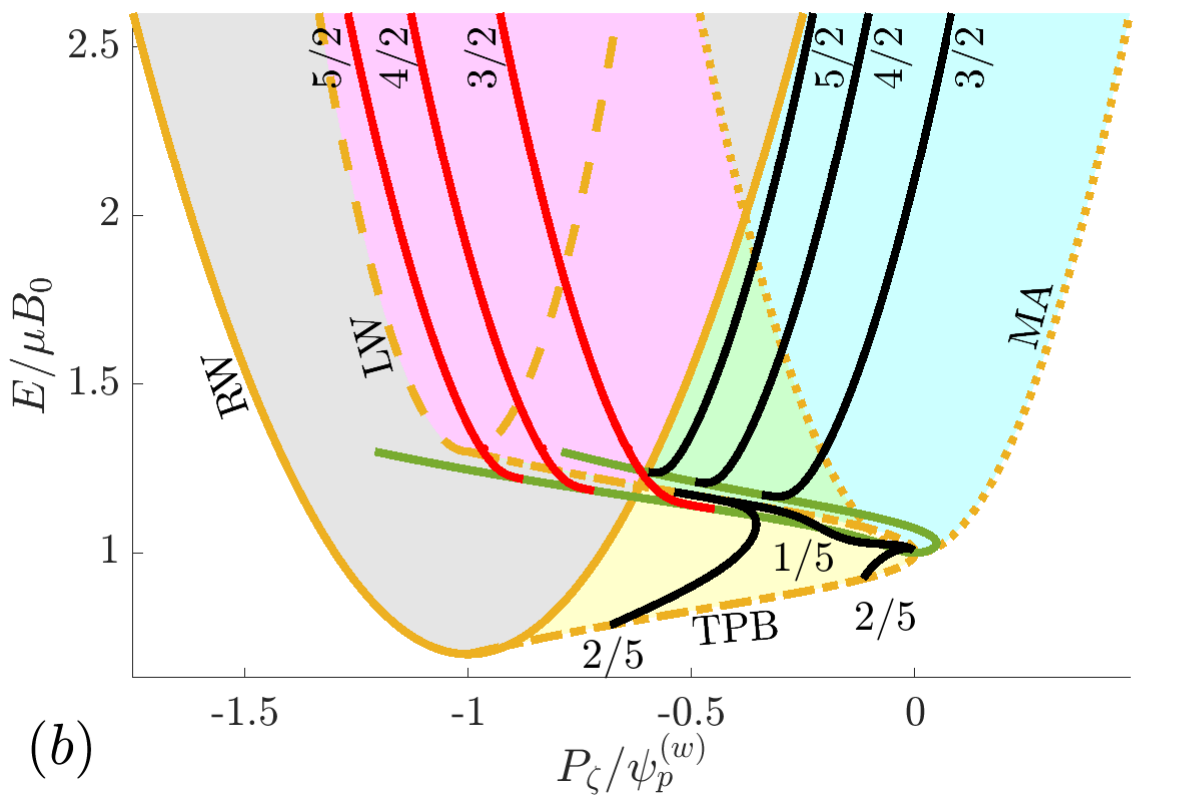}
    \includegraphics[width=0.49\textwidth,keepaspectratio]{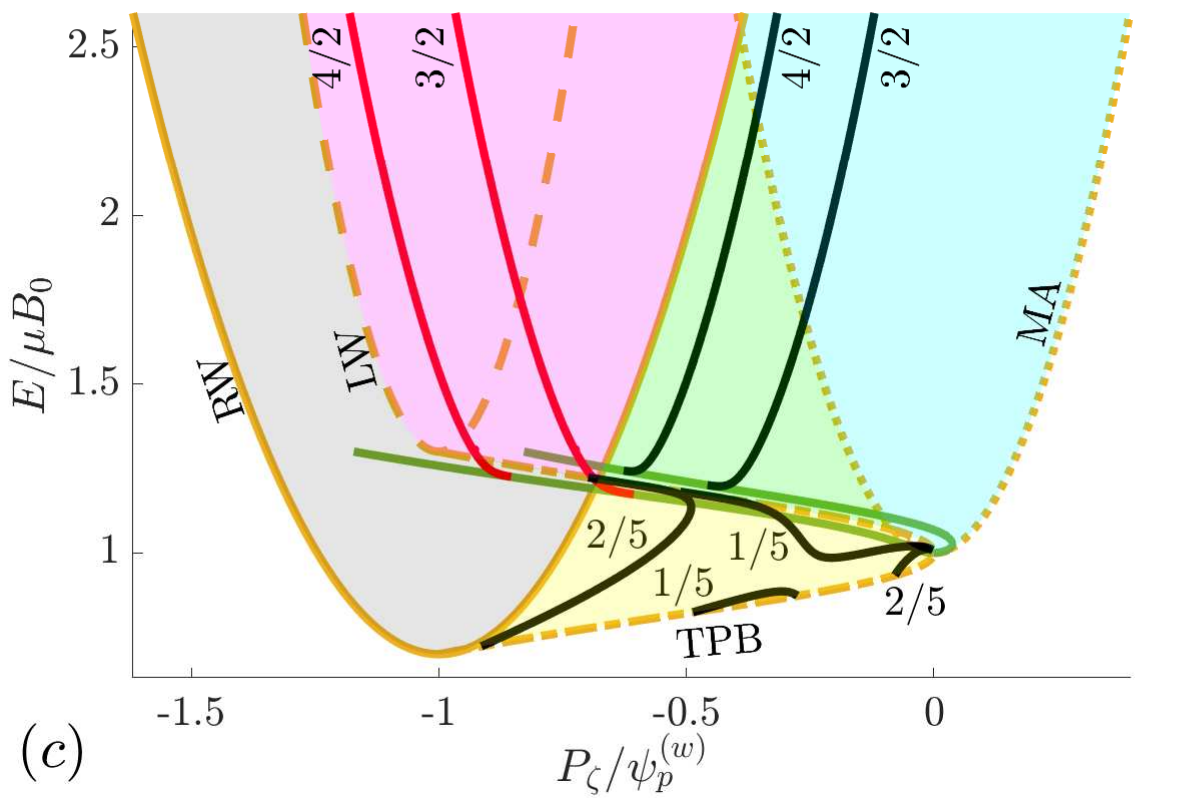}
    \includegraphics[width=0.49\textwidth,keepaspectratio]{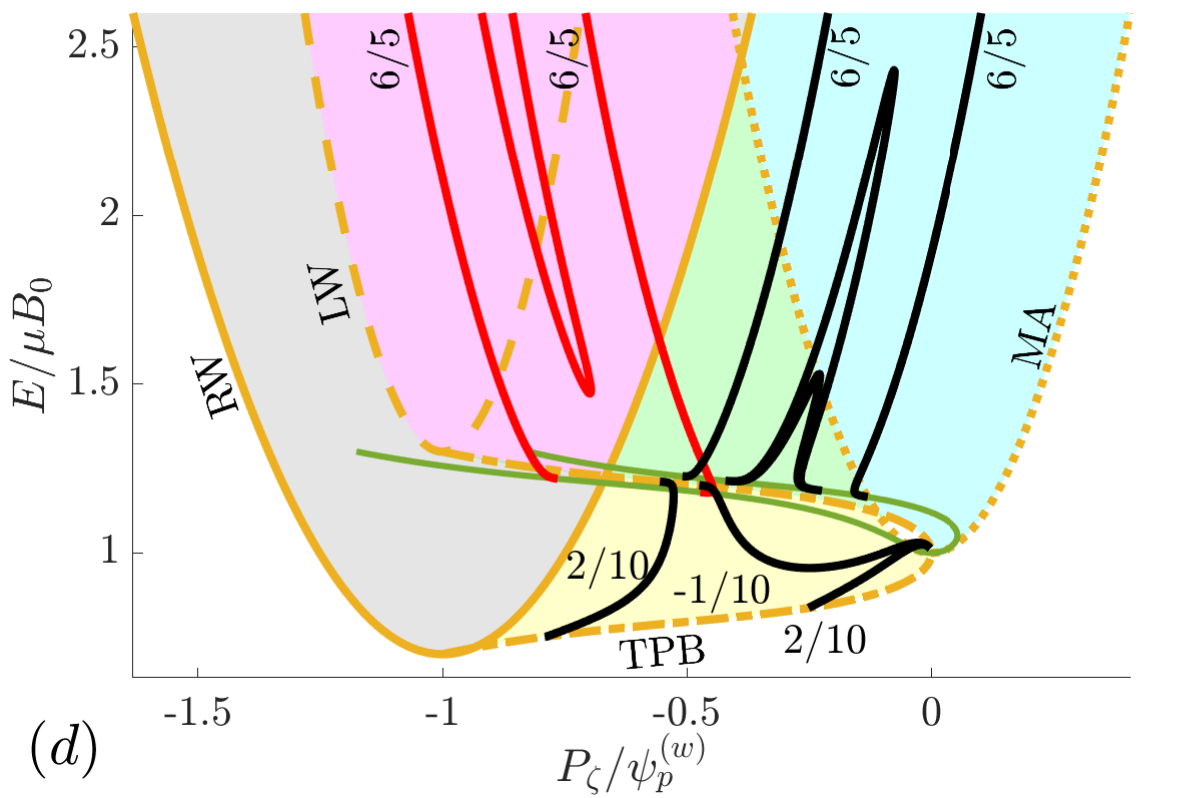}
    \caption{Constant-$\mu$ slices (plane cuts) of the three dimensional COM space $(E,\mu, P_{\zeta})$. Cases \#1-\#4 are depicted in (a)-(d), respectively. Yellow-colored parabolas depict the location of the magnetic axis (MA) (dotted line), the trapped-passing boundary (TPB) (dot-dashed line), the left wall (LW) (long-dashed line) and the right wall (RW) (plain line).  The color-shaded areas denote regions of trapped (yellow), co-passing (cyan), counter-passing (magenta), both co- and counter-passing (green), and lost (grey), particles. Black lines correspond to resonances with trapped and co-passing particles, and red lines correspond to resonances with counter-passing particles. Co-passing and counter-passing resonant curves are terminated on the green lines close to the TPB. A constant pitch $\Lambda=E/\mu B_{0}$ line intersects twice a passing particle resonance only for the Case \#4, whereas twice intersected trapped particle resonances are very close to the magnetic axis in all cases.}
    \label{fig:Fig4}
\end{figure}

A systematic assessment of the range of validity of the analytical results, based on the width of the drift GC orbits, can be provided in the space of the three COM where the boundaries of the magnetic axis and the walls are given in terms of the kinetic characteristics of the particles (Appendix A). Figure \ref{fig:Fig4} depicts characteristic curves representing the left (inner) and the right (outer) walls, and the magnetic axis, in a constant$-\mu$ plane of the three dimensional COM space $(E,\mu,P_\zeta)$ for the Cases \#1-\#4, based on their definitions in Eqs. \eqref{walls} and \eqref{magnetic axis}. For constant pitch $\Lambda=E/\mu B_0$, the distance $\Delta P_\zeta$ between the neighboring legs of the parabolas corresponding to the magnetic axis and the right wall can be used as a measure for the size of the drift width of the GC orbits (see also Sec. 3.3 of [\citenum{WhiteBook}]). Utilizing the expressions of Eqs. \eqref{walls}, \eqref{magnetic axis}, along with a first order Taylor expansion with respect to $\sqrt{2 \psi_w}$ this distance is calculated as
\begin{equation}
    \frac{\Delta P_\zeta}{\psi_{p}^{(w)}}\equiv\frac{P_\zeta^{(a)}-P_\zeta^{(rw)}}{\psi_{p}^{(w)}}=1-\frac{1+2(\Lambda-1)}{\sqrt{2(\Lambda-1)}}\left(\frac{\Delta B}{B_0}\right)\left(\frac{\sqrt{\mu}}{\psi_{p}^{(w)}}\right), \label{DPzeta} 
\end{equation}
where $P_\zeta^{(a)}, P_\zeta^{(rw)}$ are the values of $P_\zeta$ at the magnetic axis and the right wall, respectively, and $\Delta B / B_0 = \sqrt{2 \psi_w}=r$ ($r$ is the minor radius) is the variation of the magnetic field amplitude from the magnetic axis to the walls. This equation clearly shows the role of the kinetic characteristics, along with the value of the poloidal flux at the wall $\psi_{p}^{(w)}$, on the the drift orbit width and the corresponding validity of the analytical calculations: for a given pitch $\Lambda$, $\Delta P_\zeta$ decreases with increasing $\mu$ and increases with increasing $\psi_{p}^{(w)}$. This dependence can be confirmed by observing the horizontal distance between the parabolas corresponding to the right wall and the magnetic axis in Fig. \ref{fig:Fig4}, and explains the validity of the analytical results and its dependence on the kinetic characteristics of the particles and on the $q$ profile. More specifically, the comparison between Figs. \ref{fig:Fig4}(b) and \ref{fig:Fig4}(c) clearly shows the role of the $\psi_{p}^{(w)}$, leading to the conclusion that a higher value of the total poloidal flux (higher plasma current) results in smaller particle drifts.

The employment of a flux surface of reference $\psi_0$ for the magnetic field evaluation, in order to obtain the analytical expressions for the $q_{kin}$, also influences the approximation of the trapped-passing boundary. In the absence of such approximation, the trapped-passing boundary is uniquely defined by Eq. \eqref{trapped passing boundary}. However, under the above approximation, the trapped-passing boundary is defined by the condition $k=1$ or $E = \mu(1+\sqrt{2P_{\theta_{0}}})$ which, by the definition of $P_{\theta_{0}}$, as in Eq. \eqref{def of Ptheta0},  gives different values for trapped, co-passing and counter-passing particles. More specifically, for trapped particles the approximate condition aligns with the exact trapped-passing boundary, whereas for co- and counter-passing particles, the respective condition gives two different trapped-passing boundaries  (yellow dot-dashed and green lines, respectively, in Fig. \ref{fig:Fig4}). Notably, the approximate curves for co-passing and counter-passing particles are located above and bellow the exact trapped-passing boundary, respectively.   

Having calculated the kinetic $q$ factor as a function of the three COM, its level sets at rational values $q_{kin}=m/n$ can be readily obtained. Several curves corresponding to different rational values are depicted in Fig. \ref{fig:Fig4}, for trapped (black), co-passing (black) and counter-passing (red) particles. These curves show the exact locations in the COM space for the particles that can resonantly interact with specific non-axisymmetric perturbations, as will be shown in the following section, and provide an overview for the resonant response of the toroidal plasma. Moreover, it is worth noting that, in accordance to the previous discussion referring to the extrema of the $q_{kin}$, only for the non-monotonic $q$ profile of Case \#4, a constant pitch $\Lambda=E/\mu B_{0}$ line intersects twice a passing particle resonance, whereas twice intersected trapped particle resonances are very close to the magnetic axis, in all cases.     

\section{Resonant response to non-axisymmetric perturbations} \label{Resonse}
Non-axisymmetric time-independent perturbative magnetic modes can be described as
\begin{equation}
    \delta \boldsymbol{B} = \nabla \times \alpha\boldsymbol{B}
\end{equation}
 where $\alpha$ is defined as 
 \begin{equation} \label{alpha}
\alpha(\psi,\theta,\zeta)=\sum_{m,n} \alpha_{m,n}(\psi)e^{i(m\theta-n\zeta)}
 \end{equation}
with $m,n$ corresponding to the poloidal and toroidal mode numbers, and $\alpha_{m,n}(\psi)$ being the radial profile of the mode amplitude. This form of the magnetic field perturbations sufficiently describes the radial component of all perturbations of practical interest, including MHD activity such as toroidal field ripples or Resonant Magnetic Perturbations (RMP) \cite{White2013a,White2013b, White2018}. Moreover, this form of the magnetic perturbation can be readily incorporated into the canonical GC equations by directly modifying the normalised parallel velocity $\rho_\parallel$ to $\rho_\parallel+\alpha$ in Eq.\eqref{GC H}, resulting in the modified GC Hamiltonian \cite{White1982, White1984, White2013}
 \begin{equation} \label{Per GC H 1}
    H = \frac{\left[(P_{\zeta} + \psi_{p}(P_{\zeta},P_{\theta}) - \alpha(P_{\zeta},P_{\theta}, \zeta, \theta)\right]^2}{2 g(P_{\zeta},P_{\theta})} B^2(P_{\zeta}, P_{\theta}, \theta) + \mu B(P_{\zeta}, P_{\theta}, \theta),
\end{equation}

In order to analyze the resonant effects of the perturbations on the GC orbits with the utilization of the previously calculated kinetic $q$ factor, the expression of Eq.\eqref{alpha} has to be transformed to AA variables 
 \begin{equation} \label{alpha hat}
    \alpha(J_{\theta},J_{\zeta},\hat{\theta},\hat{\zeta})= \sum_{m',n'}{\hat{a}_{m',n'}(J_{\theta},J_{\zeta})e^{i(m'\hat{\theta} - n'\hat{\zeta})}},
\end{equation}
allowing the resonance condition to be straightforwardly expressed as 
\begin{equation}
    q_{kin}(E,\mu,P_\zeta)=\frac{m'}{n'}.
\end{equation}
It is important to emphasize that, due to the nonlinear character of the transformation from the geometrical angles $(\theta,\zeta)$ to the Angles $(\hat{\theta},\hat{\zeta})$ according to Eq.\eqref{eq:Angles}, although $n'=n$, there exists no direct correspondence between the poloidal mode numbers $m$ and $m'$. In fact, even a single perturbative mode in Eq.(26) may lead to multiple modes in Eq.(28). The impact of a single geodesic acoustic-like compressional mode on fast-ion transport and losses due to resonances overlapping at fractional harmonics of the mode frequency has been observed in DIII-D tokamak experiments \cite{Nazikian2008}. A theoretical study explaining the fractional resonances that appear under geodesic acoustic mode as well as other MHD activity, like Alfvén eigenmodes, has been presented in \cite{Kramer2012}, whereas the role of primary and sideband resonances in the confinement of energeting ions under the presence of RMPs has been considered for EAST \cite{He2020}. The relation between the two variable sets crucially determines the modification of the GC phase space due to the applied perturbations, as will be shown in the following paragraphs and further discussed in Appendix B.

\begin{figure}
    \centering
    \includegraphics[width=0.49\textwidth,keepaspectratio]{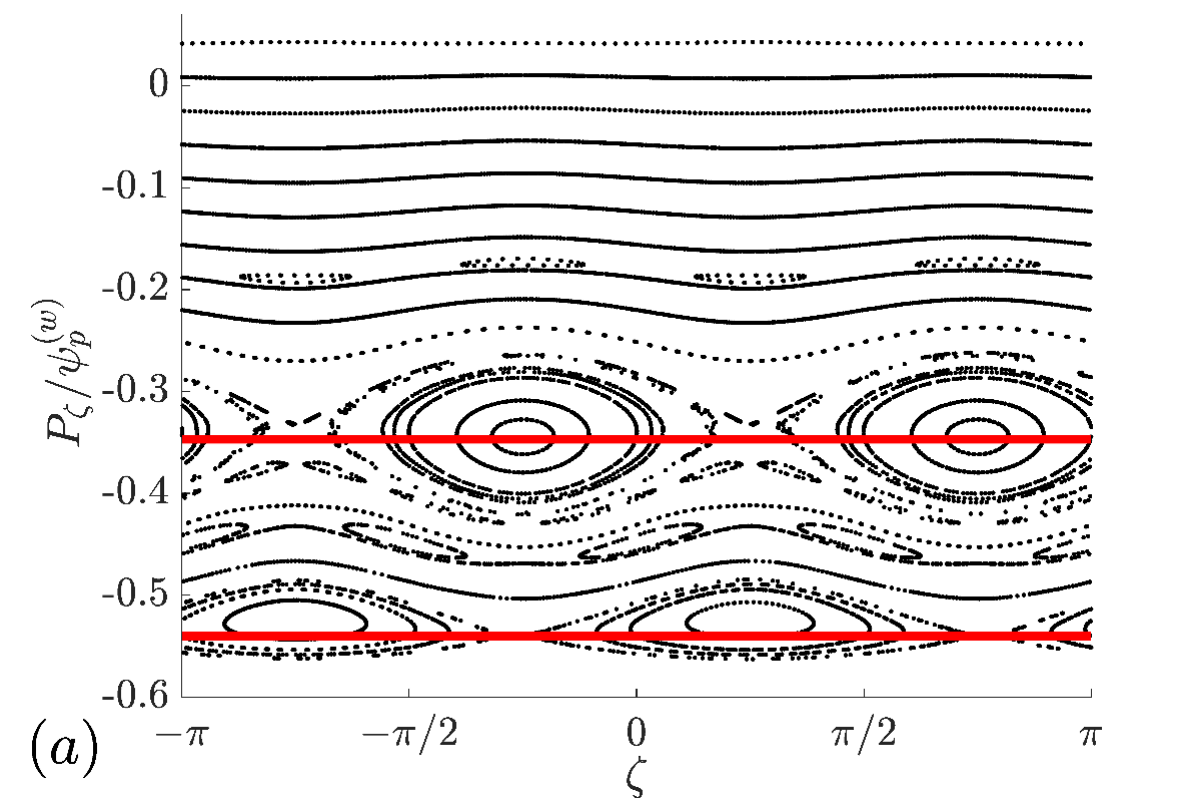}
    \includegraphics[width=0.49\textwidth,keepaspectratio]{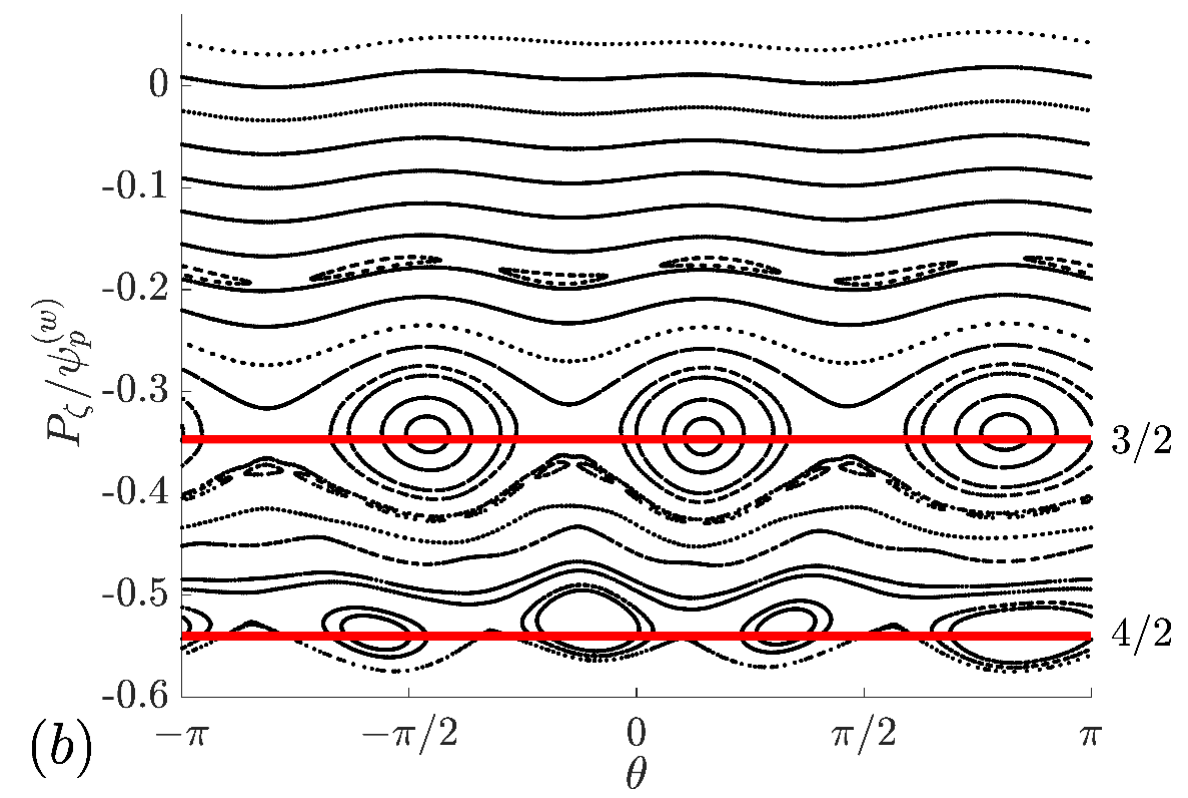}
    \includegraphics[width=0.49\textwidth,keepaspectratio]{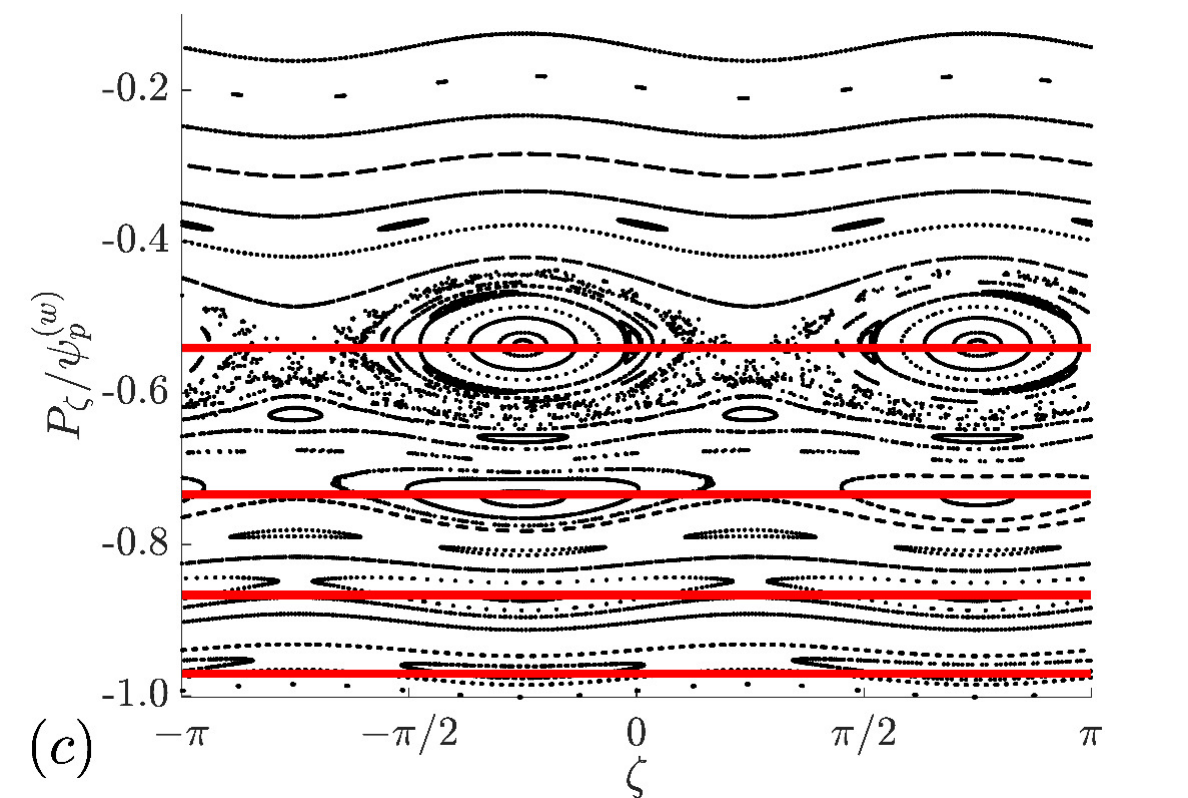}
    \includegraphics[width=0.49\textwidth,keepaspectratio]{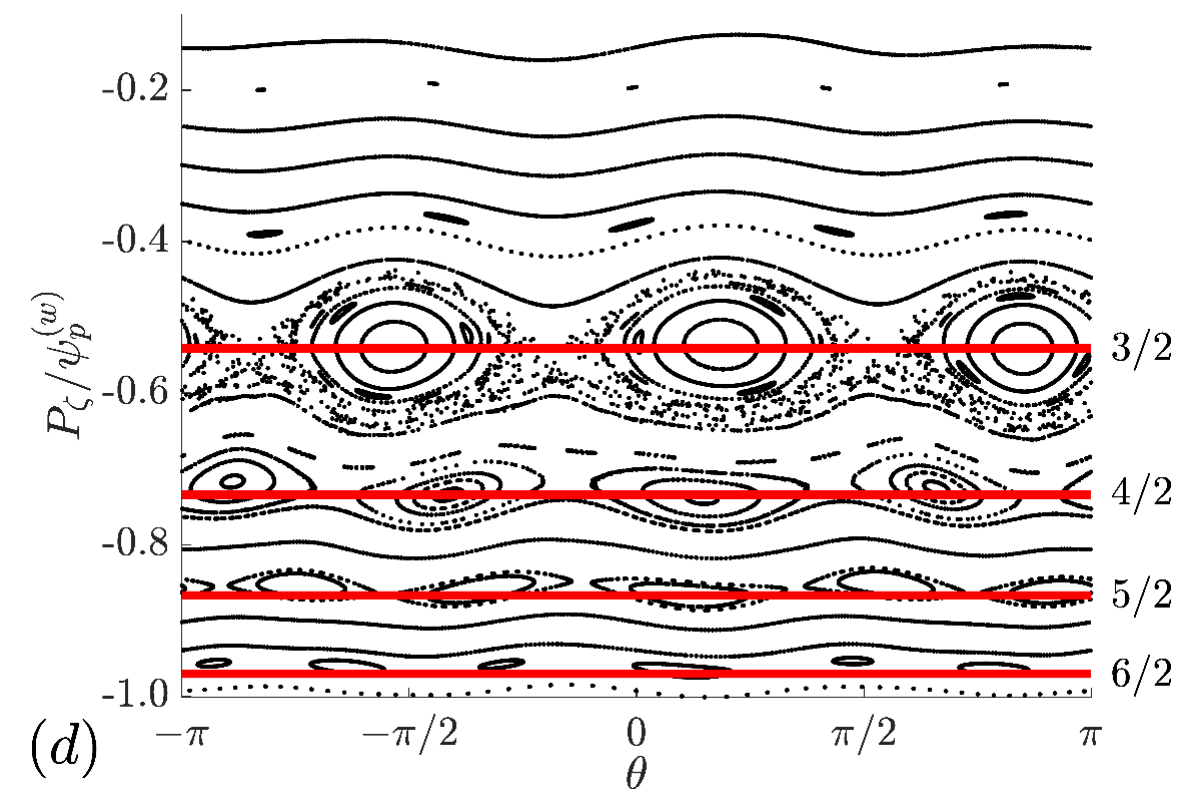}
    \caption{Poincaré surfaces of section for passing particles of Case \#1 ($\mu B_{0} = \SI{2}{\kilo\electronvolt}$ in the $q_{1}$ profile), under the presence of a single perturbative mode $(m,n)=(3,2)$ with amplitude $\alpha_{3,2}=0.5\times10^{-4}$. Co-passing particles with $E/\mu B_{0} = 1.20$, crossing the surfaces of section in the positive directions are depicted in panels (a) and (b), while counter-passing particles with $E/\mu B_{0} = 1.43$, crossing the surfaces of section in the negative direction are depicted in panels (c) and (d). Panels (a) and (c) display Poincaré surfaces of section at constant poloidal angle $\theta=0$, whereas panels (b) and (d) show Poincaré surfaces at a constant toroidal angle $\zeta=0$. The red horizontal lines denote the predicted locations of the corresponding resonant island chains in accordance to Fig. \ref{fig:Fig4}(a), showing a remarkable accuracy.}
    \label{fig:Fig5}
\end{figure}

Without loss of generality, concerning the resonant character of mode-particle interactions and their effects on stochastic transport, in the following analysis, we consider perturbations with a constant mode-amplitude, that is, $\alpha_{m,n}=\epsilon$, where $\epsilon$ represents the ratio of the amplitude of the perturbative magnetic mode to the background magnetic field. The effect of the perturbative modes on the GC phase space is strongly inhomogeneous and localized in regions where the kinetic characteristics of the particles satisfy the respective resonance conditions. In order to dissect the phase space, we employ appropriate Poincaré surfaces of section at either $\theta=0$ or $\zeta=0$, for fixed values of $\mu$ and $E$. For illustrative purposes, we focus on the Case \#1. According to Fig. \ref{fig:Fig4}(a) the pitch values of $E/\mu B_{0}=1.2$ and $E/\mu B_{0}=1.43$ correspond to passing particles with different ranges of $P_{\zeta}/\psi_{p}^{(w)}$. The Poincaré surfaces of section, under the presence of a single perturbing mode with $(m,n)=(3,2)$, are depicted in  Fig. \ref{fig:Fig5}(a)-(b) for co-passing particles with pitch value $E/\mu B_{0}=1.2$,  and Fig. \ref{fig:Fig5}(c)-(d) for counter-passing particles with pitch value $E/\mu B_{0}=1.43$. It is clear that even a single perturbing mode can generate several resonant island chains centered around values of $P_\zeta$ where $q_{kin}$ factor has a rational value. The number of islands in each island chain in the $\theta=0$ surface is equal to the toroidal mode number $n$, whereas the number of islands in each chain in the $\zeta=0$ surface is equal to the different poloidal mode numbers $m'$ resulting from the transformation to Action-Angle variables according to the discussion in the Appendix B (Fig. \ref{fig:Fig9}(b)). The analytically obtained $q_{kin}$ accurately predicts the resonance locations, indicating the paramount importance of the respective diagram shown Fig. \ref{fig:Fig4}, which provides concise information regarding the resonant response of the particles to non-axisymmetric perturbations and pinpointing the locations of the phase space where strong mode-particle interactions actually take place. 

\begin{figure}
    \centering  
    \begin{minipage}{0.49\textwidth}
      \raisebox{1cm}{  \includegraphics[width=\textwidth,keepaspectratio]{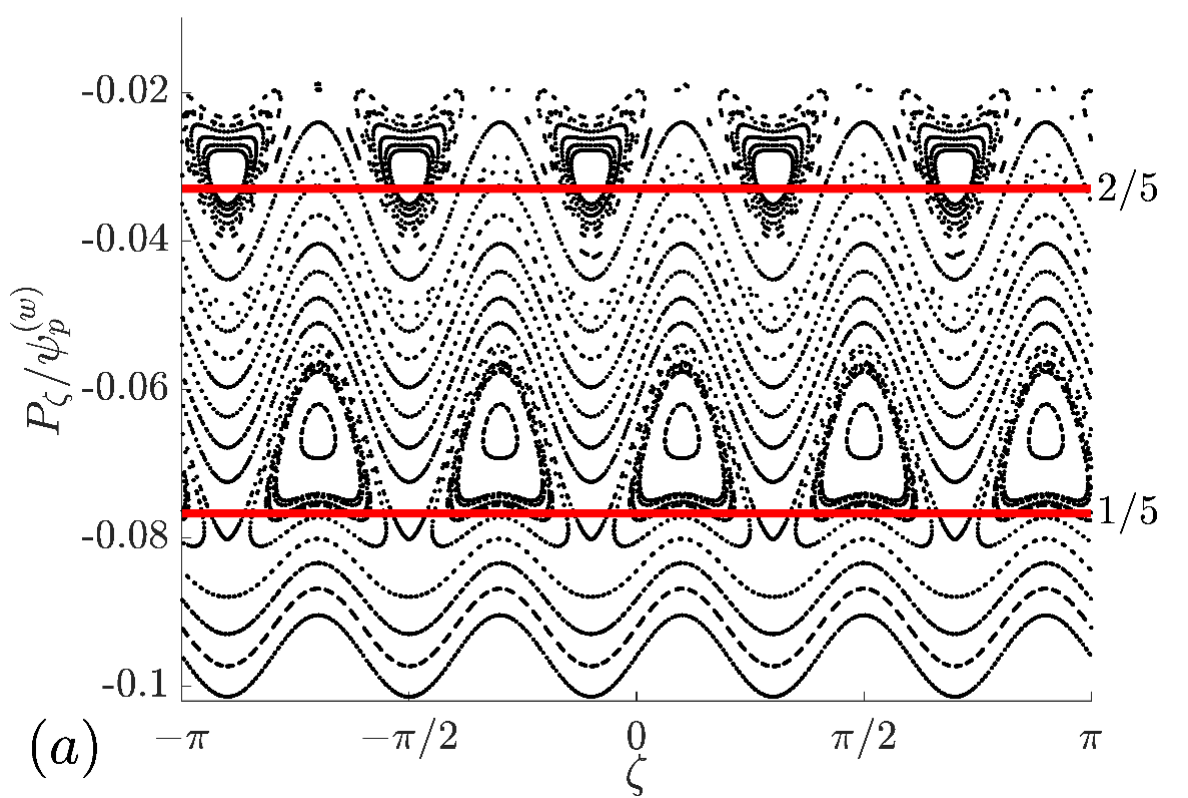}}
    \end{minipage}
    \hfill
    \begin{minipage}{0.49\textwidth}
        \includegraphics[width=\textwidth,keepaspectratio]{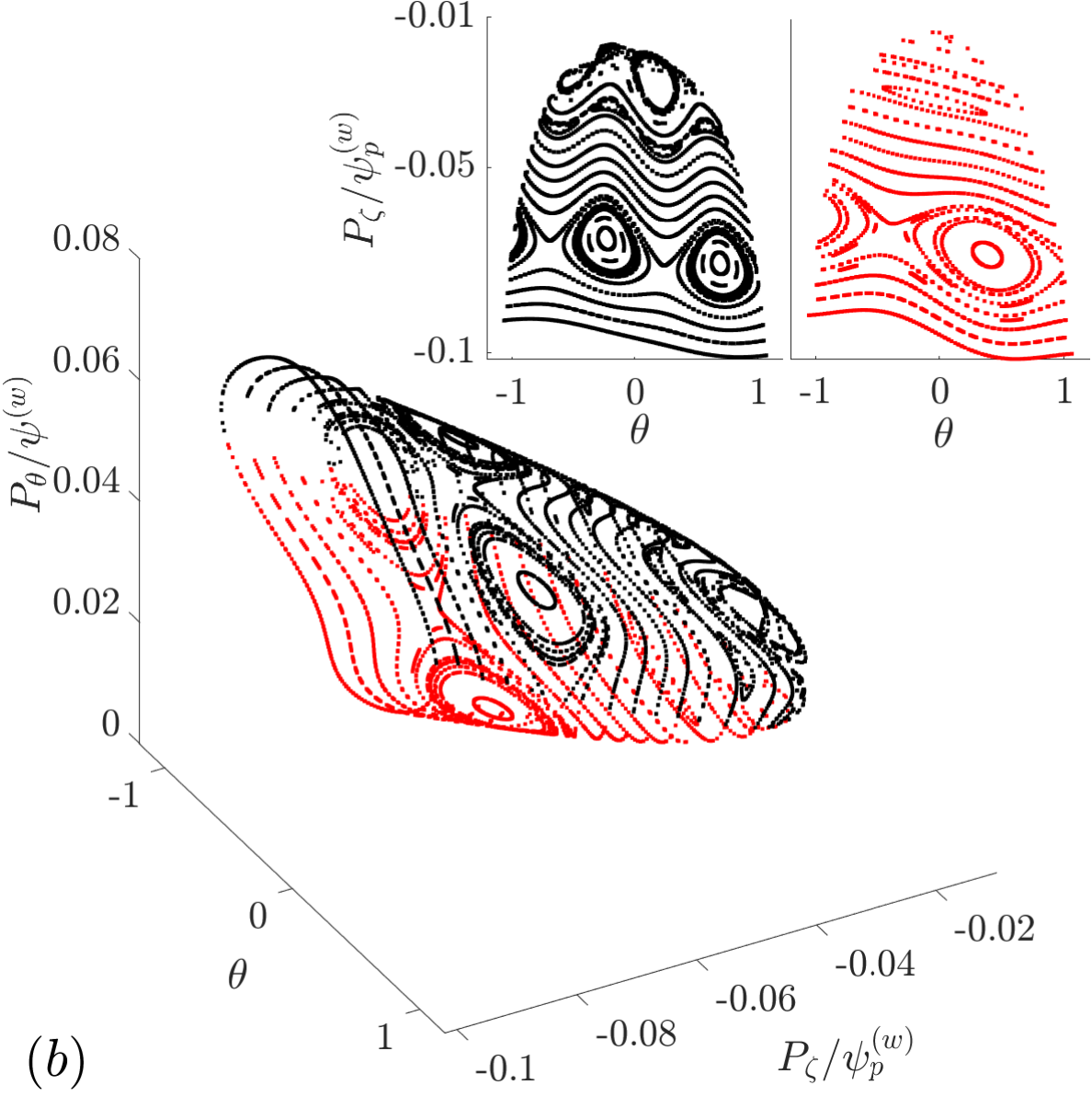}
    \end{minipage}
    \begin{minipage}{0.49\textwidth}
      \raisebox{1cm}{  \includegraphics[width=\textwidth,keepaspectratio]{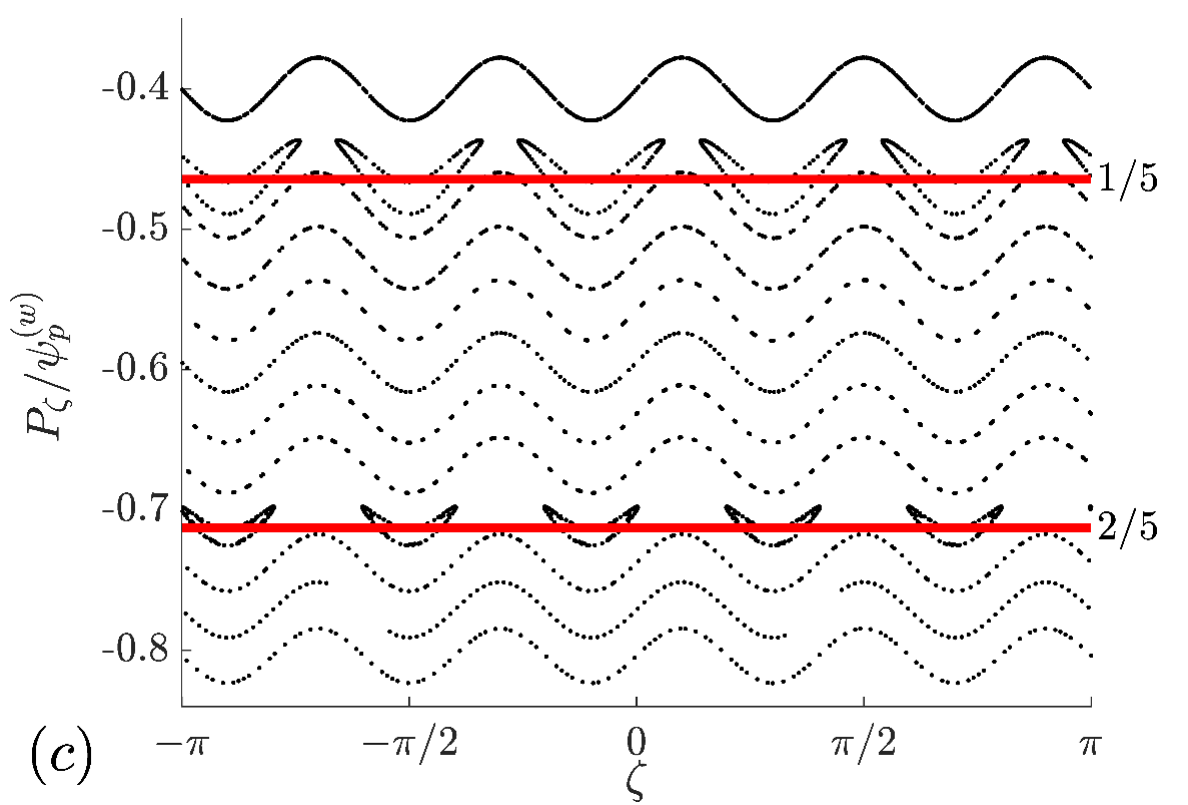}}
    \end{minipage}
    \hfill
    \begin{minipage}{0.49\textwidth}
        \includegraphics[width=\textwidth,keepaspectratio]{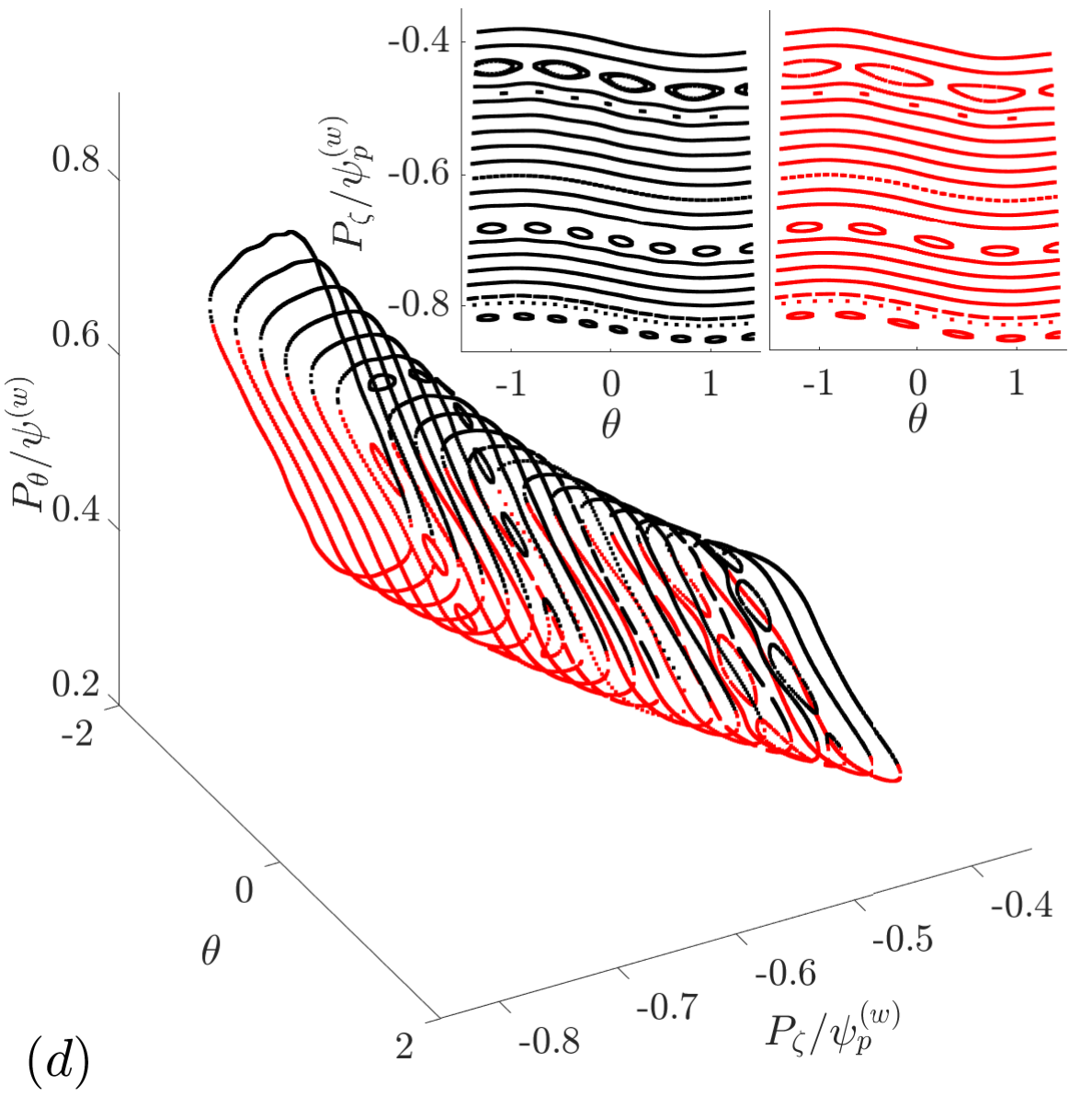}
    \end{minipage}

    \caption{Poincaré surfaces of section for trapped particles of Case \#1 ($\mu B_{0} = \SI{2}{\kilo\electronvolt}$ in the $q_{1}$ profile) with $E/\mu B_{0} = 0.996$, under the presence of two perturbative modes, $(m,n) = (1,5)$ and $(2,5)$. 
    Cases of two different perturbation strengths $\alpha_{1,5} = \alpha_{2,5} = 0.5 \times 10^{-4}$ and $\alpha_{1,5} = \alpha_{2,5} = 2.1 \times 10^{-4}$ are depicted in panels (a)-(b) and (c)-(d), respectively. 
    Panels (a) and (c) display Poincaré surfaces of section at the constant poloidal angle $\theta=0$, with red horizontal lines denoting the predicted locations of the corresponding resonant island chains in accordance to Fig. \ref{fig:Fig4}(a).
    Panels (b) and (d) show Poincaré surfaces at a constant toroidal angle $\zeta=0$, with black and red color indicating orbit intersections with the surface of section in positive and negative directions, respectively. The 3D diagrams depict intersections in both directions lying in a two-dimensional constant energy surface.} 
    \label{fig:Fig6}
\end{figure}

 A characteristic case of trapped particles with pitch $E/\mu B_{0}=0.97$ under the presence of  two perturbative modes with mode numbers $(m,n)=(1,5),(2,5)$ is depicted in Fig. \ref{fig:Fig6}. On the $\theta=0$ Poincaré surfaces of section, shown in Fig. \ref{fig:Fig6}(a) and (c), the locations of the resonances are accurately predicted by the analytical results. In accordance to Fig. \ref{fig:Fig4}(a), each resonance appears at two locations of the phase space and the number of islands in each resonance chain corresponds to the toroidal mode number $n$. The $\zeta=0$ Poincaré surfaces of section, defined by orbit intersection in the positive and negative directions are shown in Fig. \ref{fig:Fig6}(b) and \ref{fig:Fig6}(d), respectively, with the orbits appearing to be open. A clearer view of the phase space topology can be provided by a generalized Poincaré surface of section where orbit traces of both positive and negative intersections are depicted and shown to lie in a two-dimensional constant energy surface in the three-dimensional space ($P_\zeta,P_\theta,\theta$). The number of islands in each resonant chain (in both directions) can be accurately predicted from the corresponding unperturbed orbit on the basis of the discussion presented in Appendix B, as shown by comparing the number of crossings depicted in Fig. \ref{fig:Fig9}(a) with the number of islands in the $q_{kin}=2/5$ resonant chain shown in Fig. \ref{fig:Fig6}(d).

\begin{figure}
    \centering
    \includegraphics[width=0.49\textwidth,keepaspectratio]{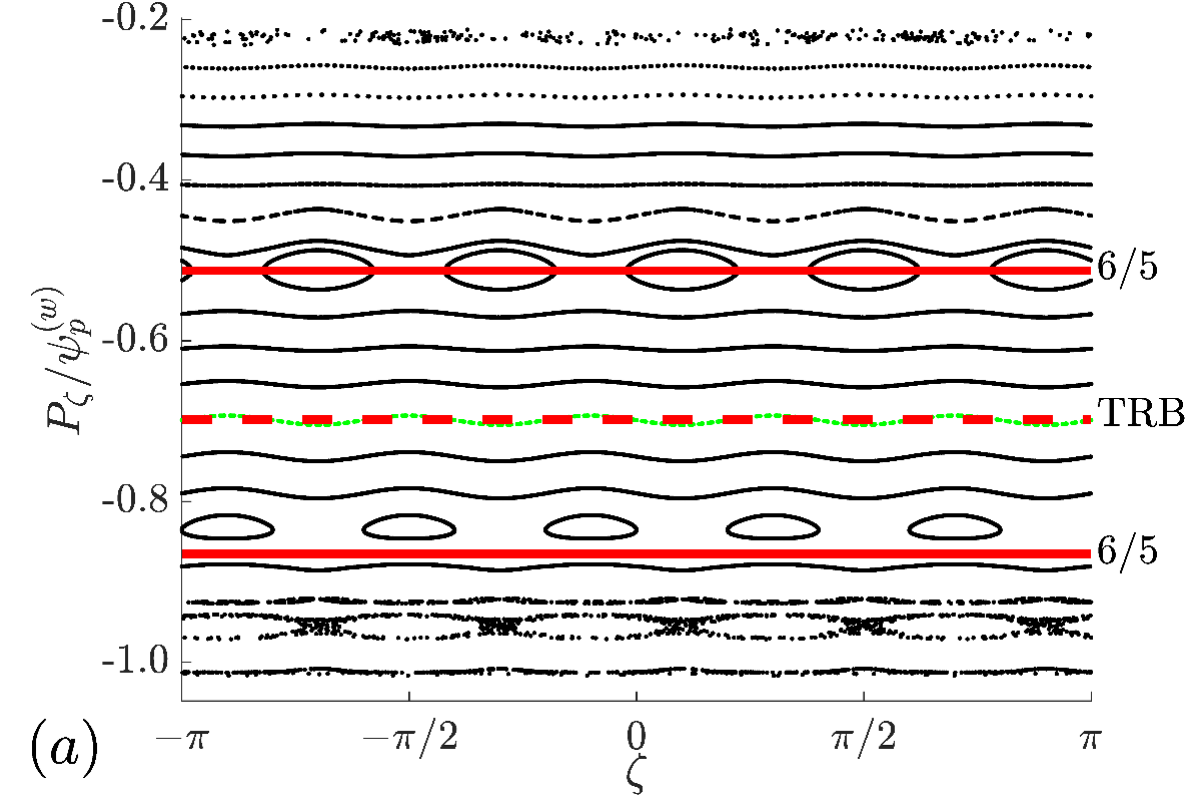}
    \includegraphics[width=0.49\textwidth,keepaspectratio]{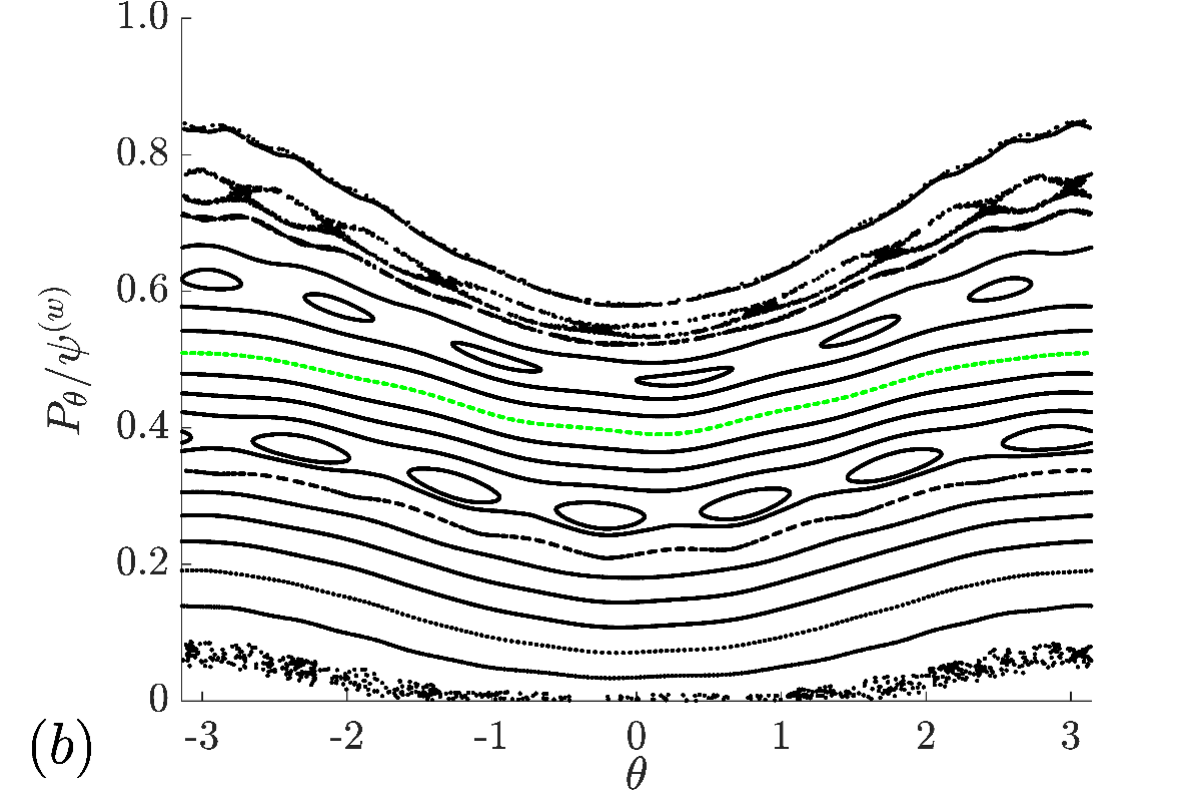}
    \includegraphics[width=0.49\textwidth,keepaspectratio]{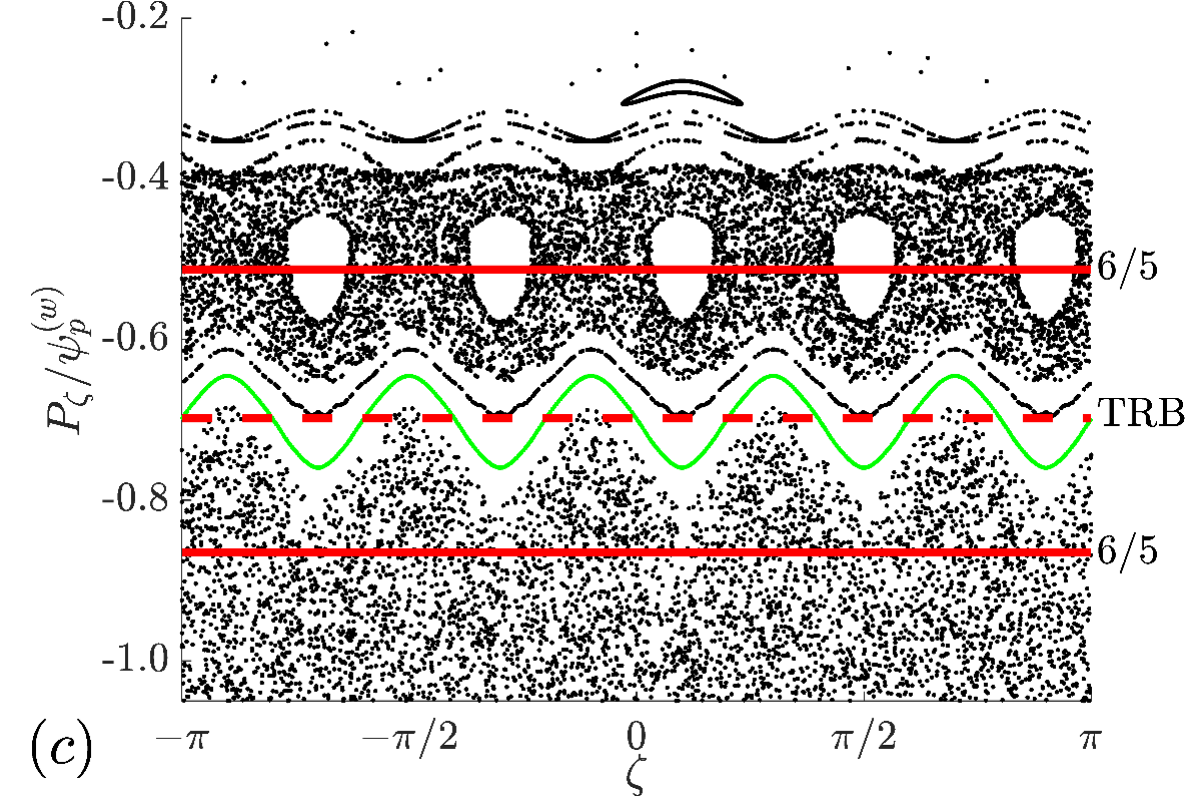}
    \includegraphics[width=0.49\textwidth,keepaspectratio]{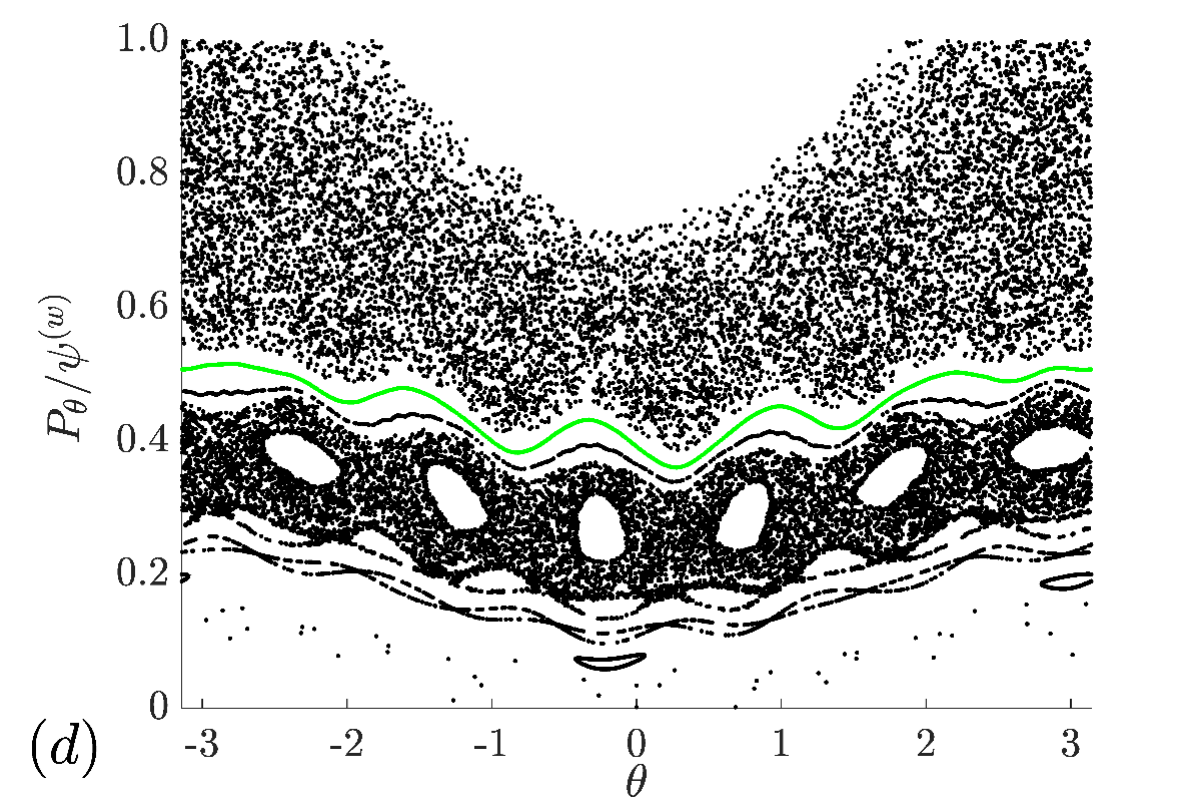}
    
    \caption{Poincaré surfaces of section for counter-passing particles of Case \#4 ($\mu B_{0} = \SI{10}{\kilo\electronvolt}$ in the $q_{3}$ profile) with $E/\mu B_{0} = 1.43$, under the presence of a single perturbative mode $(m,n)=(6,5)$. Cases of two different perturbation strengths $\alpha_{6,5}=1.1\times 10^{-5}$ and $\alpha_{6,5} = 1.1\times 10^{-4}$ are depicted in panels (a)-(b) and (c)-(d), respectively.
    Panels (a) and (c) display Poincaré surfaces of section at the constant poloidal angle $\theta=0$, whereas panels (b) and (d) show Poincaré surfaces at a constant toroidal angle $\zeta=0$. The solid red horizontal lines denote the predicted locations of the corresponding resonant island chains and the dotted red horizontal line denotes the predicted location of the transport barrier, in accordance to Fig. \ref{fig:Fig4}(a) [and Fig. \ref{fig:Fig3}(b)]. The green curves indicate the transport barriers located at a value of $P_\zeta/\psi_p^{(w)}$ corresponding to the minimum of the $q_{kin}$, shown in the mid panel of Fig. \ref{fig:Fig3}(b).} 
    \label{fig:Fig7}
\end{figure}

As mentioned in a previous section, a non-monotonic $q$ profile (Case \#4) results in the presence of a local minimum in the kinetic $q$ factor for passing particles, as shown in Fig. \ref{fig:Fig3}(b) and the corresponding appearance of certain resonances in two locations in the COM space, as depicted in Fig. \ref{fig:Fig4}(d). The Poincaré surfaces of section for a characteristic case of co-passing particles with $E/\mu B_{0} = 1.43$ is depicted in Fig. \ref{fig:Fig7}, under the presence of a single perturbative mode with mode numbers $(m,n)=(6,5)$, for two different values of the perturbation strength. In addition to the location of the resonance chains, the location of a transport barrier is predicted at the value of $P_\zeta$ that corresponds to a local minimum of the $q_{kin}$ \cite{Anastassiou2024}. The significance of the transport barrier formation becomes more evident for increasing perturbation strength, as a large part of the phase space connected to the wall has become chaotic and the transport barrier prevents particles from escaping towards the wall.

\section{Summary and Conclusions}\label{Conclusions}
A Hamiltonian Action-Angle formulation of the GC motion is utilized for the analytical calculation of the kinetic $q$ factor. The comparison between analytical and numerical results demonstrates a remarkable agreement, while the domain of validity of the respective analytical formulas is systematically investigated and explained. The calculation of the kinetic $q$ factor allows for the representation of resonance curves in the three-dimensional space of the kinetic characteristics (Constants of the Motion) of the particles, along with the curves characterizing the particles as trapped or passing, and confined or lost. These diagrams provide an overview of the plasma's kinetic response under perturbing modes and allow to accurately pinpoint the resonance island chains in the phase space at locations corresponding to rational values of the kinetic $q$ factor. Moreover, the exact number of islands within a specific resonance chain is predicted on the basis of unperturbed particle orbits with its non-trivial correspondence to the perturbing mode numbers demonstrated for the case of trapped particles. Both the locations and the number of islands of resonance chains are systematically confirmed by numerically calculated Poincaré surfaces of section. Finally, conditions for the existence of transport barriers along with their locations, corresponding to local extrema of the kinetic $q$ factor are accurately predicted and their implications for enhanced confinement are discussed.

In conclusion, the calculation of the kinetic $q$ factor, performed analytically for a LAR equilibrium in this work, is shown to enable the a priori prediction of the effect of perturbative modes in terms of the presence of resonant island chains and transport barriers in the phase space. Even for more general equilibria, this is a low-computational-cost semi-analytical calculation performed in the absence of perturbations, suggesting a valuable tool for the physical understanding and the prediction of transport and confinement properties of toroidal fusion plasmas in the presence of multi-scale perturbations.  

\section*{Appendix A: Orbit classification in the COM space for a circular LAR equilibrium}
For a circular LAR equilibrium, it is shown \cite{White1984} that $I(\psi) \simeq 0$, $g(\psi) \simeq 1$ and $\delta(\psi,\theta) = 0$. Subsequently, Eq. \eqref{canon moments} yields $P_{\theta}=\psi$ and the magnetic field can be expressed in canonical variables as $B = \left(1 - \sqrt{2 P_{\theta}}\cos{\theta}\right)$. Accordingly, the equations of the GC motion are written as
\begin{eqnarray} 
    \dot{P}_{\zeta} &=& 0 \label{dot Pzeta}\\
    \dot{\zeta} &=& \rho_{||}B^2 \label{dot zeta} \\
    \dot{P}_\theta &=&  -\frac{\partial B}{\partial \theta}\left[\rho_{||}^2 B + \mu\right] \label{dot Ptheta} \\
    \dot{\theta} &=&  \frac{1}{q(P_{\theta})}\rho_{||}B^2 + \frac{\partial B}{\partial P_{\theta}}\left[\rho_{||}^2 B + \mu\right] \label{dot theta}
\end{eqnarray}
with $\rho_\parallel=P_\zeta+\psi_p(\psi)$.

The fixed points in the poloidal projection are given by the following system of algebraic equations
\begin{eqnarray}
    -\frac{\partial B}{\partial \theta}\left[\left(P_{\zeta}+\psi_{p}\right)^2 B + \mu\right]&=&0\\
    \frac{\partial \psi_{p}}{\partial P_{\theta}}\left[P_{\zeta}+\psi_{p}(P_{\theta})\right]B^2 + \frac{\partial B}{\partial P_{\theta}}\left[\left(P_{\zeta}+\psi_{p}(P_{\theta})\right)^2 B + \mu\right]&=&0.
\end{eqnarray}
The first equation gives $\theta = 0,\pm\pi$, and, to  leading order in $\rho$, the second gives $P_\zeta =  -\psi_{p}(P_\theta)$ ($\rho_\parallel=0$). One of the fixed points ($\theta=0$) is an elliptic center and the other ($\theta=\pm\pi$) is a hyperbolic saddle. The fixed points dictate the topology of the GC motion and the number of different family orbits. The elliptic center is surrounded by trapped (banana-shaped) orbits, whereas the hyperbolic saddle defines a heteroclinic orbit connecting the fixed points $\theta=\pm \pi$ and separating trapped and co-passing (or counter-passing) orbits. In terms of the three COM the trapped-passing boundary is given as 
\begin{equation} \label{trapped passing boundary}
    E = \mu\left(1\mp\sqrt{2\psi_{p}^{-1}(-P_{\zeta})}\right)
\end{equation}
where $\psi_p^{-1}$ is the inverse function of $\psi_p(\psi)$.

Orbits can also be classified with respect to being confined or lost. For the case of a LAR configuration the loss boundary is given by 
\begin{equation}\label{walls}
    E = \frac{\left(P_{\zeta} + \psi_{p}(\psi_{w})\right)^2}{2}\left(1\mp\sqrt{2\psi_{w}}\right)^2 + \mu\left(1\mp\sqrt{2\psi_{w}}\right),  
\end{equation}
where signs $(-)$ and $(+)$ correspond to particles co-moving and counter-moving with respect to the magnetic field direction, touching at midplane the right and the left wall, respectively. Moreover, orbits passing through the magnetic axis ($\psi=0$) satisfy the equation
\begin{equation}\label{magnetic axis}
    E = \frac{P_{\zeta}^2}{2} + \mu.  
\end{equation}
Eqs. \eqref{trapped passing boundary}, \eqref{walls} and \eqref{magnetic axis} define parabolic curves in the $(E, P_{\zeta})$ plane and allow for the characterization of each orbit in a constant-$\mu$ slice of the three-dimensional COM space (Sec. 3.3 of \cite{WhiteBook}).

\section*{Appendix B: Relation between the angle variables $(\theta,\zeta)$ and $(\hat{\theta},\hat{\zeta})$} 
The Angle variables $(\hat{\theta},\hat{\zeta})$ essentially differ from the geometrical angle variables $(\theta,\zeta)$, since the former are linear functions of time, whereas the latter are in general quasi-periodic functions of time in the interval $[0, 2\pi)$. The Angle variables are phase-like $(\hat{\theta}=\hat{\omega}_\theta t + \hat{\theta}_0, \hat{\zeta}=\hat{\omega}_\zeta t + \hat{\zeta}_0)$ with the frequencies $\hat{\omega}_\theta, \hat{\omega}_\zeta$ providing the appropriate time scaling of the oscillation in each degree of freedom. The two variable sets are related through the canonical transformation Eq. \eqref{eq:Angles}, implying that $\hat{\theta}$ shares the same period with $\theta$ (although they differ as functions of time), whereas this is not the case for $\zeta$ and $\hat{\zeta}$. In general $\zeta$ can be a non-periodic function of time unless the frequencies of $\hat{\zeta}$ and $\theta$ (or $\hat{\theta}$) have a rational ratio, i.e. $\hat{\omega}_\zeta / \hat{\omega}_\theta = m' / n'$, corresponding to a resonant orbit.   

\begin{figure}
    \centering
    \includegraphics[width=0.49\textwidth,keepaspectratio]{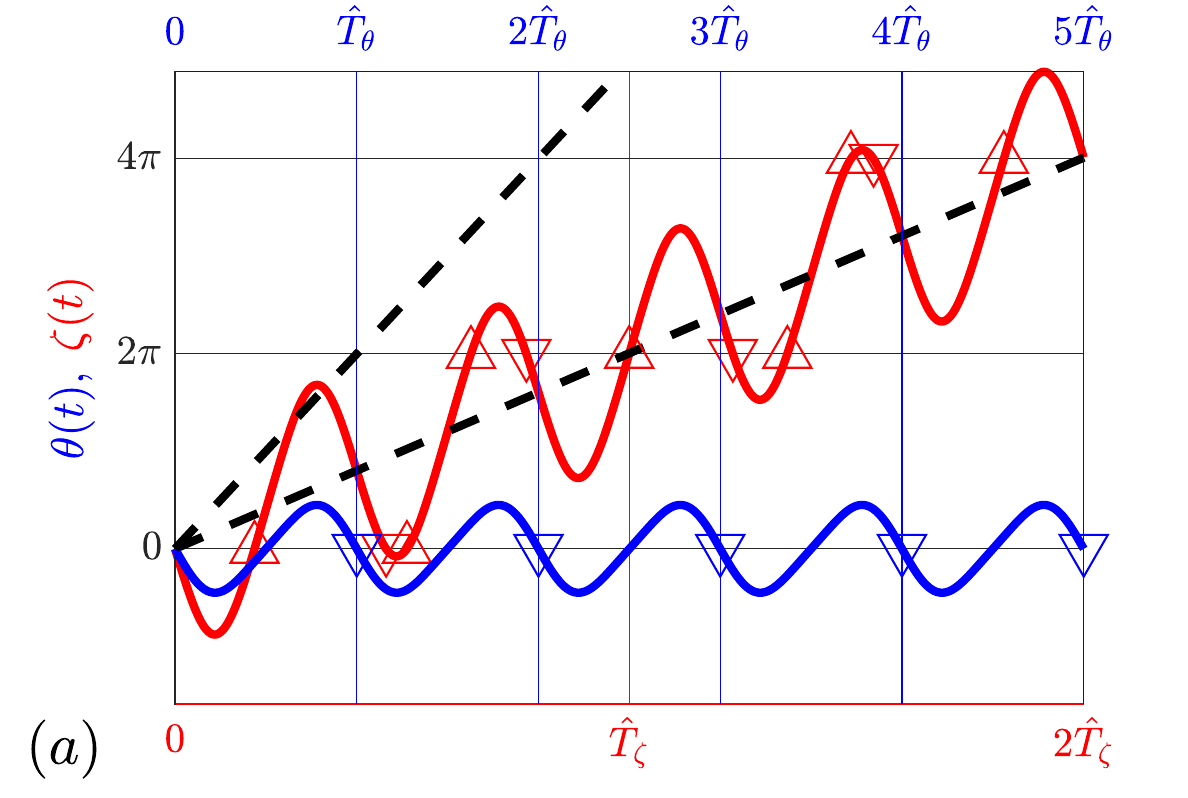}
    \includegraphics[width=0.49\textwidth,keepaspectratio]{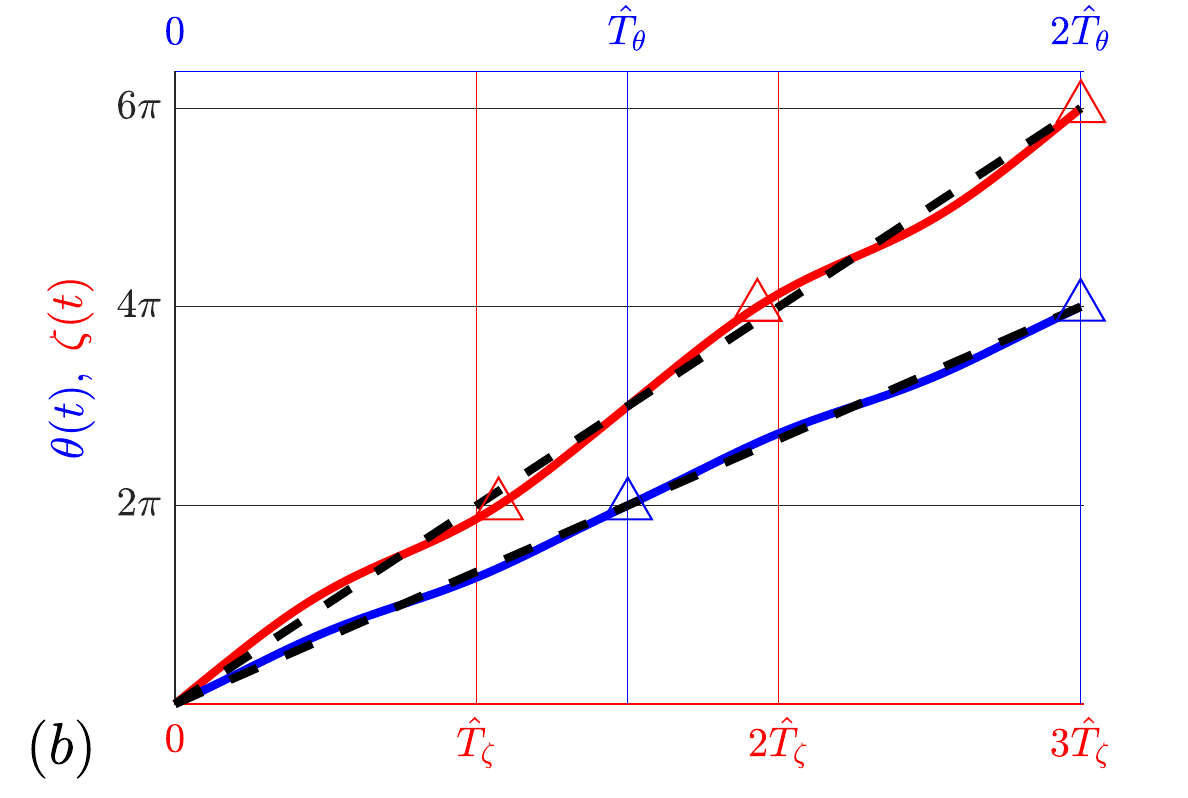}    
    \caption{Time dependence of $\theta$ (blue lines) and $\zeta$ (red lines) for a periodic trapped orbit with $q_{kin}=2/5$ (left) and a passing orbit with $q_{kin}=3/2$ (right). The time dependence, measured in $\hat{T}_\zeta$ and $\hat{T}_\theta$ directly provides the dependence on $\hat{\zeta}$ and $\hat{\theta}$, respectively. Black dashed lines depict the time variations of $\hat{\zeta}$ and $\hat{\theta}$. The traces of the periodic orbit on the Poincaré surfaces of section at $\theta = 0$ and $\zeta = 0$ are indicated by upward red and blue triangles for positive directions, and downward triangles for negative directions.}
    \label{fig:Fig8}
\end{figure}

The equations of the GC motion \eqref{dot Pzeta}-\eqref{dot theta} imply that, for passing particles $(\rho_\parallel \neq 0)$, $\dot{\zeta}(t)\neq 0$ and $\dot{\theta}(t)\neq 0$, making $\zeta(t)$ and $\theta(t)$ monotonic functions of time. Given the linear relation of the Angles with time, they are also monotonic functions of the Angles, and therefore there is a one-to-one correspondence between the two variable sets. However, this is not the case for trapped particles for which $\rho_\parallel = 0$ at the turning points (where $\dot{\zeta}(t) = 0$), and at the banana tips where ($\dot{\theta}(t) = 0$). The relation between the two variable sets, as well as their time dependence, is depicted in Fig. \ref{fig:Fig8}, both for passing and trapped particles.

From the canonical transformation Eq. \eqref{eq:Angles} it is clear that $\theta$ and $\hat{\theta}$ are related through Fourier series expansions as
\begin{eqnarray}
    \theta^{(t)}&=&\sum_k a_k(\mathbf{J})\exp(ik\hat{\theta})\\
    \theta^{(p)}&=&\hat{\theta}+\sum_l a_l(\mathbf{J})\exp(il\hat{\theta})
\end{eqnarray}
for trapped particles (libration type of motion) and passing particles (rotation type of motion), respectively \cite{Goldstein}. These equations facilitate the transformation of a non-axisymmetric perturbative mode in Action-Angle variables as follows:
\begin{equation} \label{mode}
\exp\left\{i(m\theta-n\zeta)\right\}=\exp\left\{i\left[ \left(m\theta(\hat{\theta})-n\frac{\partial f}{\partial J_\zeta}(\mathbf{J},\theta(\hat{\theta})) \right)-n\hat{\zeta}\right]  \right\}.   
\end{equation}
This clearly shows that a single $(m,n)$ mode in the original angles $(\theta,\zeta)$ results in a multitude of modes $(m',n)$ with different poloidal mode numbers $m'$ in the Angle variables $(\hat{\theta},\hat{\zeta})$.

The above analysis suggests that a single non-axisymmetric perturbative mode results, in general, in multiple resonant islands chains in different locations of the phase space where the condition $q_{kin}=m'/n$ is fulfilled, with the width of these chains depending on the corresponding mode amplitudes as expressed in Action-Angle coordinates. It is worth emphasizing that although the resonance condition expresses periodicity of the motion in the Angle variables, it also implies periodicity of the motion in the original variables, as indicated by the transformation Eq. \eqref{eq:Angles}. However, due to the nonlinear character of the transformation, the two periods might be different. Consequently, the number of cycles in the two degrees of freedom within a period may differ between the original and the Angle variable sets. 

\begin{figure}
    \centering
    \includegraphics[width=0.49\textwidth,keepaspectratio]{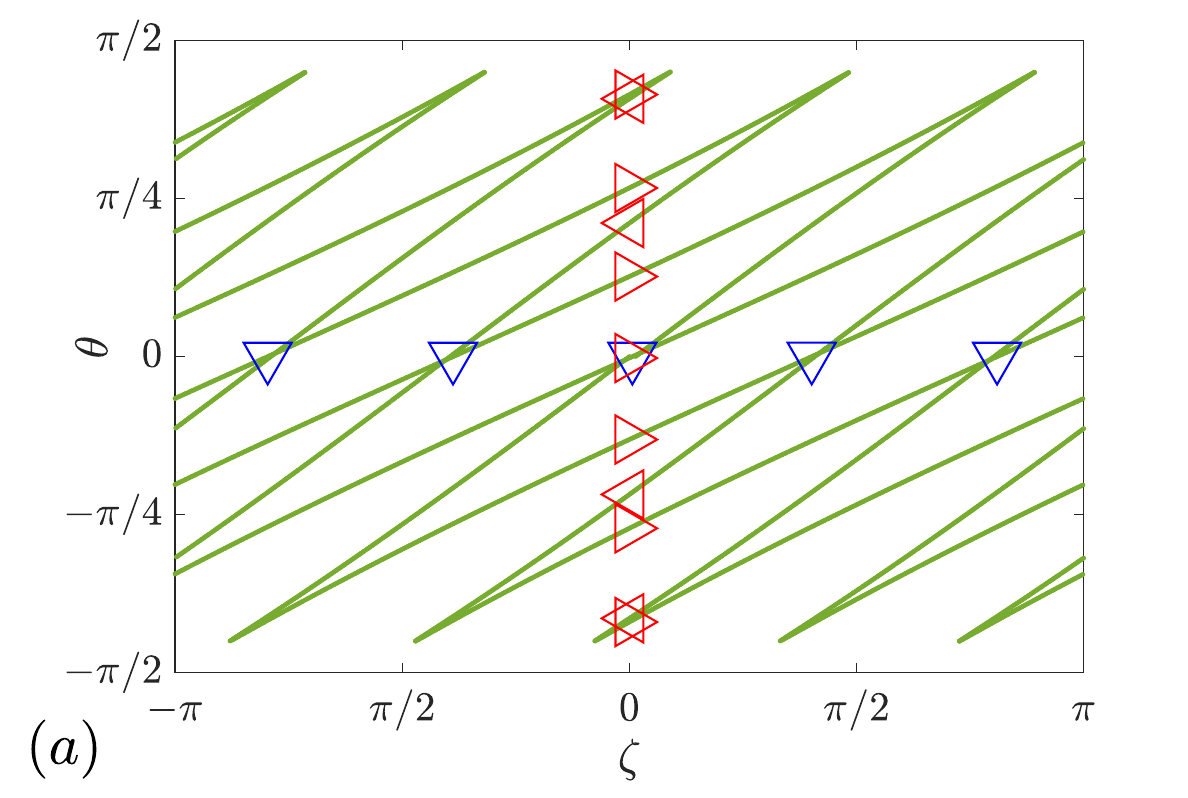}
    \includegraphics[width=0.49\textwidth,keepaspectratio]{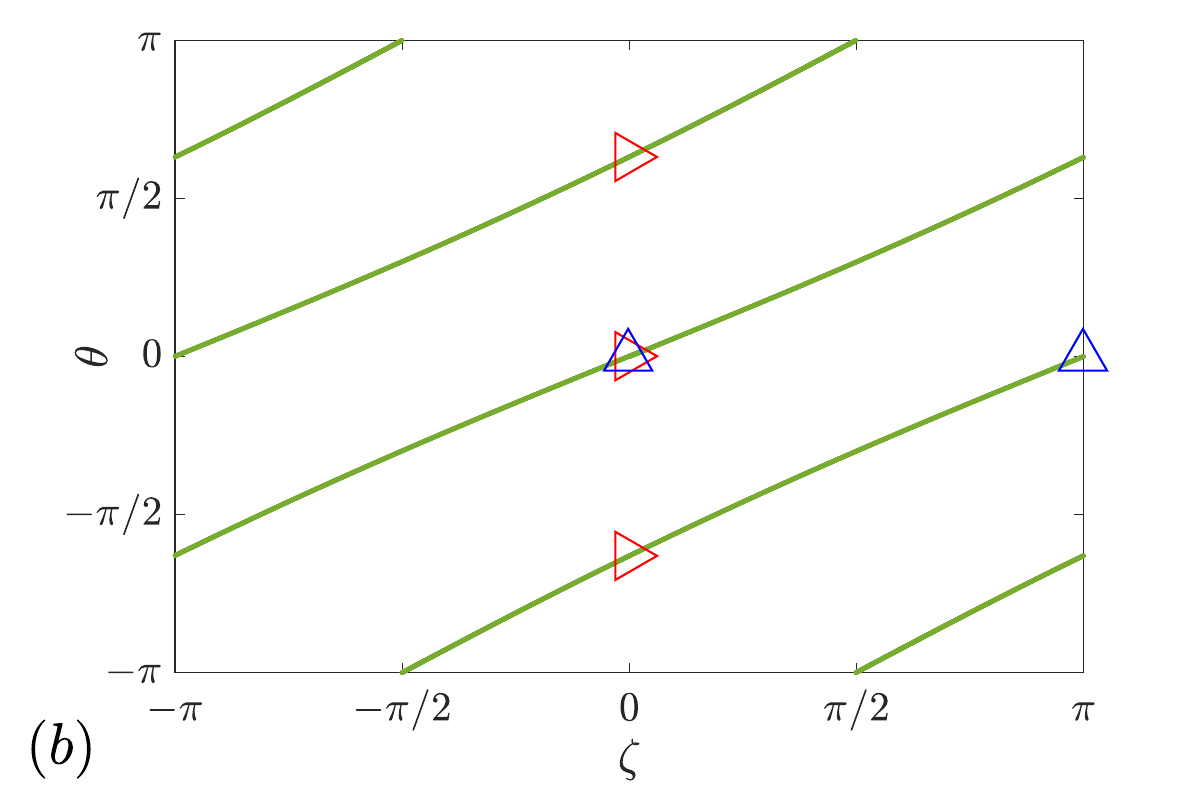}
    \caption{Winding of $\theta$ and $\zeta$ for a periodic trapped orbit with $q_{kin}=2/5$ (a) and a passing orbit with $q_{kin}=3/2$ (b). Upward and downward blue triangles indicate the traces of the periodic orbit on the $\zeta = 0$ Poincaré surface in positive and negative directions, respectively, while left- and right-facing red triangles denote the traces on the $\theta = 0$ Poincaré surface in positive and negative directions, respectively.}
    \label{fig:Fig9}
\end{figure}

Under the presence of perturbations, the  same resonance conditions apply for both variable sets, although a different number of resonances appear, due to Eq. \eqref{mode}, and a different number of islands appear within each resonance chain. The islands in a Poincaré surface of section of the perturbed system surround elliptical points corresponding to the traces of a periodic orbit on the Poincaré surface of section, and therefore their number can be a priori determined by counting these traces for an unperturbed periodic (resonant) orbit as shown in Fig. \ref{fig:Fig9}. For a passing periodic orbit with $q_{kin}=3/2$ the number of traces in the Poincaré surfaces of section $\zeta=0$ and $\theta=0$ are 3 and 2, respectively, suggesting that the numbers of islands in the resonant chain under the presence of a $(m,n)=(3,2)$ perturbative mode correspond to these mode numbers, and correctly predicting the number of islands in the Poincaré surfaces of section depicted in Figs. \ref{fig:Fig5}(a) and \ref{fig:Fig5}(b). However, for a trapped orbit with $q_{kin}=2/5$ the number of traces in the Poincaré surfaces of section $\zeta=0$ and $\theta=0$ are 11 and 5, respectively, suggesting that under the presence of a $(m,n)=(2,5)$ perturbative mode the number of islands in the $\zeta=0$ Poincaré surface of section will differ from the mode number, and predicting the number of islands depicted in Figs. \ref{fig:Fig6}(c) and \ref{fig:Fig6}(d).

\section{\label{Acknowledgments} Acknowledgments}
This work has been carried out within the framework of the EUROfusion Consortium, funded by the European Union via the Euratom Research and Training Programme (Grant Agreement No 101052200 — EUROfusion). Views and opinions expressed are however those of the author(s) only and do not necessarily reflect those of the European Union or the European Commission. Neither the European Union nor the European Commission can be held responsible for them. The work has also been partially supported by the National Fusion Programme of the Hellenic Republic – General Secretariat for Research and Innovation. 

\bibliography{AIP_arxiv}

\end{document}